%% file: SmartCommunitySimulator_ArXiv_Paper_v1.tex
\documentclass[conference]{IEEEtran}
\IEEEoverridecommandlockouts
\usepackage{cite}
\usepackage{amsmath,amssymb,amsfonts}
\usepackage{algorithmic}
\usepackage{graphicx}
\usepackage{textcomp}
\usepackage{xcolor}
\def\BibTeX{{\rm B\kern-.05em{\sc i\kern-.025em b}\kern-.08em
    T\kern-.1667em\lower.7ex\hbox{E}\kern-.125emX}}

\usepackage{graphicx,wrapfig}
\usepackage{subcaption}
\usepackage{url}
\usepackage{accents}
\newcommand{\ubar}[1]{\underaccent{\bar}{#1}}
\usepackage{algorithm}
\usepackage{algorithmic}
\graphicspath{{./Digest/Images/}}

\begin{document}


\title{Smart Residential Community Simulator for Developing and Benchmarking Energy Management Systems
}



\author{\IEEEauthorblockN{Ninad Gaikwad, and Anamika Dubey}
\IEEEauthorblockA{\textit{School of Electrical Engineering \& Computer Science} \\
\textit{Washington State University}\\
Pullman, USA \\
ninad.gaikwad@wsu.edu, and anamika.dubey@wsu.edu}
}


\maketitle


\begin{abstract}
Home Energy Management Systems (HEMS) are being actively developed for both individual houses and communities to support demand response in on-grid operation, and ensure resilience during off-grid scenarios. However, most simulators used for closed-loop HEMS testing are tailored to a specific distributed energy resource (DER) configuration with a fixed number of houses, limiting flexibility and scalability. This leads to additional development efforts to support diverse DER configurations across any number of houses and to integrate appropriate weather and load data pipelines. To address these limitations, we present a scalable simulator capable of modeling any number of houses in both on-grid and off-grid modes as a Gymnasium environment. Each house can have a unique DER configuration—Rooftop Solar Photvoltaics (PV)+Battery, Battery-only, PV-only, or no DER—and includes models for air-conditioning and eight grouped circuit-level loads. The simulator integrates National Solar Radiation Database (NSRDB) weather and Pecan Street load datasets, supports three default controllers (two for off-grid, and one for on-grid scenarios), and includes performance metrics and visualization tools. We demonstrate its flexibility through simulations on individual houses and a four-house community with heterogeneous DERs, benchmarking the controllers across built-in metrics, and computation time. The results highlight the simulator’s capability to systematically evaluate control policy performance under varying system configurations.

\end{abstract}





\section{Introduction}\label{section:Introduction}
As the electric grid is modernizing towards greater sustainability and resilience, buildings are playing an important role in this transition~\cite{aguero2017modernizing}. Buildings are no longer passive consumers of electricity but are becoming active participants through the increased adoption of DERs such as PV, residential batteries,  and electric vehicles (EVs)~\cite{DOE2025DER}. At the same time, the ongoing electrification of heating and transportation is contributing to rising peak demand and placing additional stress on distribution grid infrastructure~\cite{blonsky2019potential}; while unplanned outages due to extreme weather are becoming common~\cite{ren2021analysis}. Although DERs along with Heating Ventilation and Air Conditioning (HVAC) systems in a building offer substantial opportunities for flexibility~\cite{mariano2021review}, management of peak-time grid stress, and resilience under outages; their effective utilization rests on the development and deployment of intelligent energy management systems that can coordinate, and optimize DER operations in real-time~\cite{rathor2020energy}.

A substantial body of research has explored the development of energy management systems for both commercial and residential buildings, see~\cite{mariano2021review} and~\cite{leitao2020survey} respectively. With the growth of smart grid technologies, Home Energy Management Systems (HEMS) are becoming essential for helping households use energy more efficiently and interact intelligently with the grid~\cite{zhou2016smart}. For residential applications, a wide range of model-based optimization and reinforcement learning (RL) methods have been proposed to coordinate DERs and flexible loads for objectives such as cost minimization, peak shaving, and outage resilience under both grid-connected and off-grid scenarios, see~\cite{Beaudin2015home} and~\cite{shareef2018review}. However, the simulation environments used in these works are often custom-built for a specific combination of houses and DERs, and are not readily scalable to other configurations. Moreover, modeling assumptions vary significantly across studies, performance metrics are inconsistent, and the integration of weather and load data pipelines differs. This lack of standardization poses a major barrier to reproducibility, limits the ability to test a given control scheme across multiple scenarios, and hinders benchmarking of competing HEMS strategies.

Several commercial simulation platforms, such as HOMER~\cite{homer2023}, Home Energy Simulator~\cite{eutechHESim}, and the Homes Zero Net Energy Toolkit~\cite{hnzt2021}, offer capabilities for simulation, analysis and control development of various house–DER configurations. However, these tools are proprietary and closed-source, limiting their accessibility for research, reproducibility, and customization. In the open-source domain, EnergyPlus~\cite{energyplus} and the ecosystem of tools developed around it~\cite{simulationresearchLBL} offer detailed modeling and control capabilities, primarily for single-building simulation and control development. However, their focus on individual buildings limits scalability to multi-home scenarios, and the steep learning curve hinders accessibility for rapid prototyping and testing.

To the best of our knowledge, the open-source platform that most closely aligns with our objectives is CityLearn~\cite{nweye2025citylearn}, which enables RL-based control of energy storage and HVAC systems in an urban setting. However, CityLearn does not support control of household-level loads or PV generation, and its HVAC model is tailored to commercial settings, where thermal control is continuous. In contrast, residential HVAC systems often operate under binary or hysteresis-based control strategies. Moreover, CityLearn does not consider the important off-grid operational constraint of HVAC startup power requirements, which are critical for accurate modeling of PV/battery-backed standalone systems.

To address these limitations, we present the Smart Community Simulator—a scalable and modular open-source tool designed for the development, testing, and benchmarking of HEMS control strategies at both single-home and community levels. The simulator allows for arbitrary configurations of residential buildings with different DER combinations, including PV, battery, both, or neither (all DERS and loads are controllable),  and supports operation in both grid-connected and off-grid modes (implements HVAC startup power constraint). Each home includes HVAC and circuit-level grouped loads, enabling fine-grained control and realistic demand modeling. To simplify data integration, accelerate experimentation, and improve reproducibility, the simulator includes built-in pipelines for incorporating weather data from the NSRDB and residential load profiles from the Pecan Street dataset. Three default controllers—two for off-grid and one for grid-connected mode—are included, along with performance metrics and visualization capabilities. This flexible simulation environment is well-suited for the rapid prototyping of rule-based, optimization-based, and data-driven HEMS strategies.

The remainder of the paper is organized as follows: Section~\ref{section:SmartResidentialCommunitySimulator} introduces the simulator architecture and house configurations. Section~\ref{section:Modeling} describes the device models and the on-grid and off-grid physics employed in the simulator. Section~\ref{section:DefaultControllers} presents the default controllers present in the simulator. Section~\ref{section:CaseStudy} describes the setup for the case study and discusses the results. Finally, Section~\ref{section:Conclusion} provides conclusion and future work directions.

\section{Smart Residential Community Simulator}\label{section:SmartResidentialCommunitySimulator}
\begin{figure*}[ht]
    \centering
    \begin{subfigure}[t]{0.45\linewidth}
        \centering
        \includegraphics[width=\linewidth]{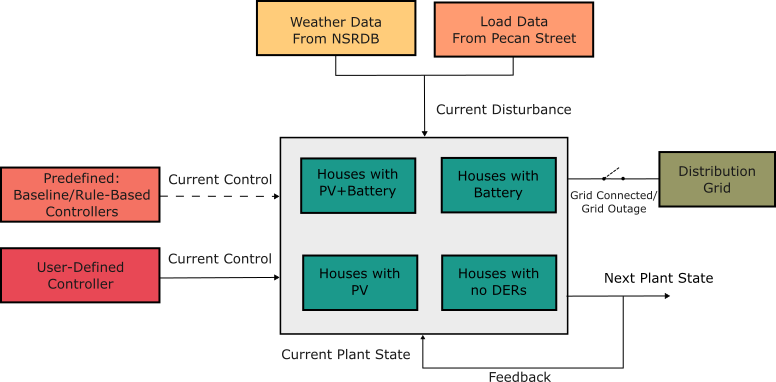}
        \caption{Schematic of Smart Community Simulator architecture.}
        \label{fig:community_houses_schematic}
    \end{subfigure}
    \hfill
    \begin{subfigure}[t]{0.45\linewidth}
        \centering
        \includegraphics[width=\linewidth]{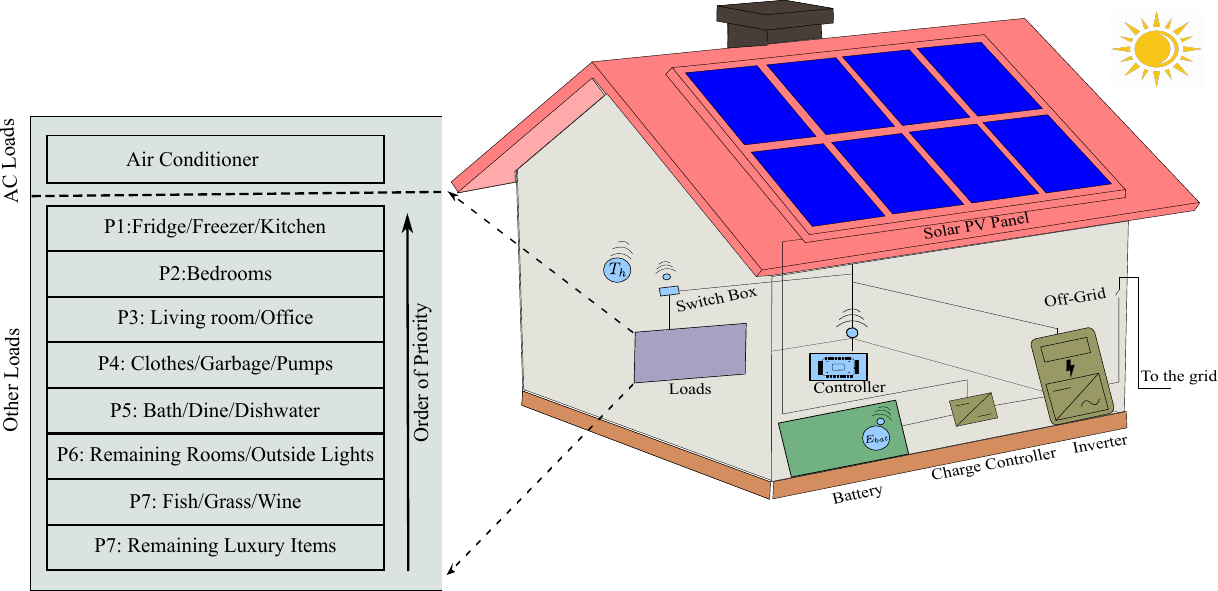}
        \caption{Schematic of a House with PV and Battery and all the loads.}
        \label{fig:community_houses_procedure}
    \end{subfigure}

    \caption{Illustration of the Smart Community Simulator architecture and configuration of a single house with PV and Battery.}
    \label{fig:community_houses_overview}
\end{figure*}

The Smart Community Simulator models residential communities with an arbitrary number of houses equipped with heterogeneous distributed energy resources (DERs). As illustrated in Fig.~\ref{fig:community_houses_schematic}, the simulator incorporates weather data from NSRDB and load profiles from the Pecan Street dataset as external disturbances. It supports both default and user-defined controllers and enables dynamic simulation of house-grid interactions under on-grid and off-grid modes. This modular setup facilitates the development and testing of intelligent control strategies for demand response, DER coordination, and resilience under realistic operating conditions.

Each house can be configured with a solar PV system, a battery storage system, or both, as shown in Fig.~\ref{fig:community_houses_procedure}. Household loads are categorized into AC and non-AC loads, with the latter further divided into eight circuit-level prioritized groups (P1–P8) based on their criticality during off-grid operation. A control policy manages PV generation, battery usage, and load operations to maintain efficient and resilient energy usage. The simulator supports four house configurations: PV+Bat, Bat-only, PV-only, and No DER.

The simulator is available as the open-source \texttt{SmartCommunitySim} package on GitHub~\cite{SmartCommunitySim}. Developed primarily in MATLAB with a Python wrapper, it provides a Gymnasium-compatible~\cite{towers2024gymnasium} environment for simulation, analysis, and intelligent control design. It includes tools for preprocessing weather and load data and comes with a curated library of NSRDB and Pecan Street datasets in both raw and processed formats to support flexible simulation and control design workflows.

\section{Modeling}\label{section:Modeling}
Time is considered to be discrete, indexed by $k = 0, 1, 2, 3, \dots$, with $\Delta T_s$ denoting the time step. The variable $E(k)$ represents the energy produced or consumed by any subsystem during the interval between time steps $k-1$ and $k$. When appropriate, the dependence on $k$ may be omitted for clarity, and constants will be explicitly indicated. Variables marked with an overbar ($\bar{.}$) or underbar ($\ubar{.}$) denote constants or known data. The units used for energy, power, and temperature are $kWh$, $kW$, and $^{\circ}C$, respectively.

Let $\mathcal{H}$ be the set of all houses in a community. We define four disjoint subsets of $\mathcal{H}$: $\mathcal{H}_{PV+Bat}$, $\mathcal{H}_{Bat}$, $\mathcal{H}_{PV}$, and $\mathcal{H}_{None}$, representing houses with both PV and battery, only battery, only PV, and no DERs, respectively. Each house $i \in \mathcal{H}$ belongs to exactly one of these subsets.

\subsection{House Thermal Model}\label{subsection:HouseThermalModel}
The thermal dynamics of each house are modeled using a discrete-time approximation of a $4^{\text{th}}$-order ODE model from \cite{cui2019hybrid}. For simplicity, we represent this using a reduced first-order model. For a house $i \in \mathcal{H}$, the house temperature $T_i^h(k)$ evolves as:
\begin{align}
    T_i^h(k+1) = A T_i^h(k) + u_{i}^{ac}(k) u_{i}^{h}(k) Q_{ac} + D T_{am}(k) \; .
\end{align}
Here, $A$, and $D$ are discrete-time system constants. The binary control input $u_i^{ac} \in \{0,1\}$ represents the on/off status of the HVAC system, and $u_i^{h} \in \{-1,+1\}$ represents the heating/cooling mode of the AC. The ambient temperature is denoted by $T_{am}$. The thermal power delivered by the HVAC system is given by $Q_{ac} = COP \cdot P^{ac}_{R}$, where $COP$ is the coefficient of performance and $P^{ac}_{R}$ is the rated power of the AC unit.

\subsection{PV Model}\label{subsection:PVModel}
The PV energy output for house $i \in \mathcal{H}_{PV+Bat} \cup \mathcal{H}_{PV}$, accounting for control action $u_i^{pv}(k) \in [0,1]$ is adapted from ~\cite{tanaka2012optimal}, and is given by:
\begin{align}
    E_i^{pv}(k) &= u_i^{pv}(k) \cdot \bar{E}_i^{pv}(k) \; \\
    \bar{E}_i^{pv}(k) &=  N_i^{pv} \, P_R^{pv} \left( \frac{G(k)}{G_{std}} \right) \nonumber \\
    &\quad \times \left( 1 + \frac{\gamma}{100} \left( T_m^{pv}(k) - T_{std} \right) \right) \Delta T_s \; ,  \\
    T_m^{pv}(k) &= T_{am}(k) + \frac{G(k)}{U_0 + U_1 + W_s(k)} \; .\label{eq:PV_Model}
\end{align}
Here, $E_i^{pv}$ and  $\bar{E}_i^{pv}(k)$ are the controllable PV energy output and potential energy output for house $i$ respectively, $N_i^{pv}$ is the number of PV panels, and $P_R^{pv}$ is the rated power of each panel. $\gamma$ is the temperature coefficient of power, $T_m^{pv}$ is the PV module temperature, and $T_{std}$, $G_{std}$ denote standard test conditions. $G$, $T_{am}$, and $W_s$ represent solar irradiance, ambient temperature, and wind speed, respectively (weather disturbance). The module temperature is computed using Faiman’s model~\cite{faiman2008assessing}.

\subsection{Battery Model}\label{subsection:BatteryModel}
Battery dynamics are modeled as a bucket of energy. For any house $i \in \mathcal{H}_{PV+Bat} \cup \mathcal{H}_{Bat}$, the battery dynamics are given by:
\begin{align}
    & E_i^{bat}(k+1) = E_i^{bat}(k) + \eta_{bat}^c \, E_i^{bat,c}(k) + \frac{E_i^{bat,d}(k)}{\eta_{bat}^{d}} \label{eq:battery_energy}  ,\\
    & E_i^{bat,c}(k) = c_i(k)  \min\left\{ \frac{\bar{E}^{bat} - E_i^{bat}(k)}{\eta_{bat}^c}, \bar{E}^{bat,c}, E_{i}^{m} \right\}  , \label{eq:battery_charge} \\
    & E_i^{bat,d}(k) = d_i(k)  \min\left\{ \left( E_i^{bat}(k) - \ubar{E}^{bat} \right)  \eta_{bat}^{d}, \bar{E}^{bat,d}, E_{i}^{m} \right\}  , \label{eq:battery_discharge} \\
    & E_{i}^{m}(k) = | E_{i}^{pv}(k) - E_i^{d_{2}}(k) |  .\label{eq:battery_mismatch}
\end{align}
Here, $E_i^{bat}$ is the battery energy level, $E_i^{bat,c}$ and $E_i^{bat,d}$ are the charging and discharging energies, and $\eta_{bat}^c$, $\eta_{bat}^{d}$ are the battery charging and discharging efficiencies. The control commands $c_i \in [ 0,1]$ and $d_i \in [ 0,1]$ determine the charging and discharging decisions. Battery energy is constrained within $[\ubar{E}^{bat}, \bar{E}^{bat}]$, and charging/discharging amounts are bounded by $\bar{E}^{bat,c}$ and $\bar{E}^{bat,d}$, respectively. $E_{i}^{m}$ is the energy mismatch between PV generation and total load demand ($E_i^{d_{2}}$ explained in Section~\ref{subsection:BaselineControllersOffOnGrid})

\subsection{Load Models}\label{subsection:LoadModel}
The total household load is divided into two categories: the air-conditioning (AC) load and a set of other loads indexed by priority levels. For any house $i \in \mathcal{H}$ and priority level $j \in P = \{P1, \dots, P8\}$, the load models are given by:
\begin{align}
    E_i^{ac}(k) &= u_i^{ac}(k) \cdot \bar{E}^{ac}, \label{eq:ac_load} \; ,\\
    E_i^{l,j}(k) &= u_i^{l,j}(k) \cdot \bar{E}^{l,j}_{i} \; .\label{eq:priority_load}
\end{align}
Here, $E_i^{ac}$ is the energy consumed by the AC load and $\bar{E}^{ac} = P_R^{ac} \cdot \Delta T_s$ is the rated AC energy consumption, and $E_i^{l,j}$ is the energy consumed by the $j^{th}$ priority load, with $u_i^{l,j} \in \{0,1\}$ being the control command for that load and $\bar{E}^{l,j}_{i} $ is the desired load demand (load disturbance) and the total other load demand served and desired are then $E_{i}^{l} = \sum_{P} E_i^{l,j}$ and $\bar{E}_{i}^{l} = \sum_{P} \bar{E}_i^{l,j}$ respectively. 

The AC startup power exceeds its rated power and is given by $\bar{P}^{ac}_{su} = (1 - \alpha_V) \cdot \alpha_I \cdot P_{ac}^{rated}$, where $\alpha_V = 0.3$ is the startup voltage factor (typically constant across AC systems) and $\alpha_I \in \{3, 4, 5, 6, 7, 8\}$ is the startup current factor (locked-rotor current) that varies with the AC system type \cite{AC:LRA}. This gives rise to two simulation modes: With AC Startup Constraints (WACSC) and Without AC Startup Constraints (WOACSC).

\subsection{Off-Grid and On-Grid Mode Physics}\label{subsection:OffOnGridPhysics}
The Algorithm~\ref{alg:OnOffGridPhysics} illustrates how Off-Grid and On-Grid physics is implemented. The total energy balance in the community is captured through the following equations:
\begin{align}
    E_{mis}^{com}(k) &= E_{g}^{com}(k) - E_{d}^{com}(k) \; , \label{eq:energy_mismatch} \\
    E_{grid}^{com}(k) &= E_{mis}^{com}(k) \; , \label{eq:grid_exchange} \\
    E_{g}^{com}(k) &= \sum_{i \in \mathcal{H}}  \left( E_i^{pv}(k) +  E_i^{bat,d}(k) \right) \; ,  \label{eq:total_generation} \\
    E_{d}^{com}(k) &= \sum_{i \in \mathcal{H}} \left( E_i^{ac}(k) +  E_i^{bat,c}(k) + \sum_{j \in P} E_i^{l,j}(k) \right) \label{eq:total_demand} \; .
\end{align}
Here, $E_{g}^{com}$ and $E_{d}^{com}$ denote the total energy generated and demanded in the community, respectively. $E_{mis}^{com}$ is the net mismatch between generation and demand, and $E_{grid}^{com}$ represents the net energy exchanged with the grid. Positive $E_{grid}^{com}$ values correspond to export to the grid, while negative values indicate grid import.

In Off-grid mode with WOACSC, loads are served only if the community-level energy mismatch is non-negative, i.e., $E_{mis}^{com} \geq 0$, ensuring local generation suffices; if $E_{mis}^{com} < 0$, no loads are supplied. In On-grid mode, all loads are always served, as any deficit ($E_{mis}^{com} < 0$) is met by grid import.
\begin{align}
    P_{mis}^{ac}(k) &= \frac{E_{g}^{com}(k)}{\Delta T_s} - \sum_{i \in \mathcal{H}} \delta_i^{ac}(k) \cdot \bar{P}_{su}^{ac}   \; ,\label{eq:ac_startup_mismatch} \\
    \delta_i^{ac}(k) &= 
    \begin{cases}
        1,  \text{if } u_i^{ac}(k) = 1 \text{ and } u_i^{ac}(k-1) = 0 \\
        0,  \text{otherwise}
    \end{cases} \label{eq:delta_ac} \; .
\end{align}

In WACSC mode, the AC Startup constraint ensures that the available power generation in the community is sufficient to support the increased power requirements of turning on ACs. To model this, a binary indicator $\delta_i^{ac}$ is defined for each house $i \in \mathcal{H}$, as shown in (\ref{eq:delta_ac}). Using this indicator, the community-level power mismatch due to AC startup is computed as in (\ref{eq:ac_startup_mismatch}), where $P_{mis}^{ac}$ denotes the AC startup mismatch. In Off-grid mode, due to the limited generation capacity of local DERs, loads are served only if $P_{mis}^{ac} \geq 0$. This constraint is not enforced in On-grid mode, where sufficient power is assumed to be always available from the grid to support AC startup.

\begin{algorithm}
\caption{Off-Grid and On-Grid Physics}
\label{alg:OnOffGridPhysics}
\begin{algorithmic}[1]
\STATE Compute total generation $E_g^{com}(k)$ using \eqref{eq:total_generation}
\STATE Compute total demand $E_d^{com}(k)$ using \eqref{eq:total_demand}
\STATE Compute energy mismatch $E_{mis}^{com}(k)$ using \eqref{eq:energy_mismatch}
\STATE Compute grid exchange $E_{grid}^{com}(k)$ using \eqref{eq:grid_exchange}

\IF{Off-Grid Mode}
    \IF{WOACSC (Without AC Startup Constraint)}
        \IF{$E_{mis}^{com}(k) \geq 0$}
            \STATE Serve all desired loads (AC and non-AC)
            \STATE Operate all DERs according to control policy 
        \ELSE
            \STATE No loads are served
            \STATE No DERs are operated
        \ENDIF
    \ELSIF{WACSC (With AC Startup Constraint)}
        \IF{$E_{mis}^{com}(k) \geq 0$}
            \FORALL{$i \in \mathcal{H}$}
                \STATE Compute $\delta_i^{ac}(k)$ using \eqref{eq:delta_ac}
            \ENDFOR
            \STATE Compute AC startup mismatch $P_{mis}^{ac}(k)$ using \eqref{eq:ac_startup_mismatch}
            \IF{$P_{mis}^{ac}(k) \geq 0$}
                \STATE Serve all desired loads (AC and non-AC)
                \STATE Operate all DERs according to control policy 
            \ELSE
                \STATE No loads are served
                \STATE No DERs are operated
            \ENDIF
        \ELSE
            \STATE No loads are served
            \STATE No DERs are operated
        \ENDIF
    \ENDIF
\ELSIF{On-Grid Mode}
    \STATE Serve all desired loads (AC and non-AC)
    \STATE Operate all DERs according to control policy 
\ENDIF

\end{algorithmic}
\end{algorithm}

\section{Default Controllers}\label{section:DefaultControllers}
We provide three default controllers for the Smart Community Simulator, which are described in the following sections.

\subsection{Baseline Controller for Off-Grid and On-Grid Modes}\label{subsection:BaselineControllersOffOnGrid}

The baseline controller, abstracts the default behavior of a commercially deployed DER system at house $i \in \mathcal{H}$, operating without centralized intelligence or supervisory coordination. It consists of four independent subsystem controllers: the AC thermostat controller, the PV controller, the battery control logic, and the load control logic. Each operates locally within its respective subsystem, assuming no smart switchboard or coordination layer is present.

\subsubsection{AC Thermostat Controller}\label{subsubsec:ACThermostatController}
The AC thermostat controller at house $i \in \mathcal{H}$ governs both the mode selection (heating or cooling) and the on/off switching logic of the air conditioning system based on the indoor temperature $T_i^h(k)$. The discrete mode variable $u_i^h(k) \in \{-1, +1\}$ selects cooling ($-1$) or heating ($+1$), while the binary control variable $u_i^{ac}(k)$ activates the cooling device. The control logic is defined as follows:
\begin{align}
u_i^h(k) &=
\begin{cases}
    -1, & \text{if } T_i^h(k) \leq \ubar{T}_i^h \; ,\\
    +1, & \text{if } T_i^h(k) \geq \bar{T}_i^h \; , \\
    u_i^h(k-1), & \text{otherwise} \; .
\end{cases}, 
\end{align}
\begin{align}
u_i^{ac}(k) &=
\begin{cases}
    1, & \text{if } T_i^h(k) \geq \bar{T}_i^{ac} \; , \\
    0, & \text{if } T_i^h(k) \leq \ubar{T}_i^{ac} \; ,\\
    u_i^{ac}(k-1), & \text{if } \ubar{T}_i^{ac} < T_i^h(k) < \bar{T}_i^{ac} \; .
\end{cases} \label{eq:ACThermostatFull}
\end{align}
This models a practical hysteresis-based thermostat with mode switching and is used identically in both Off-Grid and On-Grid modes.

\subsubsection{PV Controller Logic}\label{subsubsec:PVControllerLogic}
The PV controller at house $i \in \mathcal{H}_{PV+Bat} \cup \mathcal{H}_{PV}$ governs the use of available photovoltaic energy based on the operational mode. The PV control abstraction in Off-grid mode is given as follows:
\begin{align}
u_i^{pv}(k) &=
\begin{cases} 
\frac{E_i^{d_{1}}(k)}{\bar{E}_i^{pv}(k)}, & \text{if } \bar{E}_i^{pv}(k) \geq 0 \label{eq:PVControllerOffGrid} \; , \\
0, & \text{otherwise } \; ,
\end{cases} \\
E_i^{d_{1}}(k) &= E_i^l(k) + u_i^{ac}(k) \cdot \bar{E}^{ac} + E_i^{bat,c}(k) - E_i^{bat,d}(k) \; .
\end{align}
Here, $E_i^{d_{1}}$ is the total demand in the house $i$. In this mode, PV operates in a \textit{load-matching mode}—automatically providing energy for all active loads, including AC and battery charging. The PV controller logic in On-grid mode is given as follows:
\begin{align}
u_i^{pv}(k) =
\begin{cases}
1, & \text{if } \bar{E}_i^{pv}(k) > 0  \; , \\
0, & \text{otherwise } \; .
\end{cases} \label{eq:PVControllerOnGrid}
\end{align}
In this mode, PV is dispatcthed without curtailment, and is controllable through a supervisory controller.

\subsubsection{Battery Control Logic}\label{subsubsec:BatteryControlLogic}
The battery control logic at house $i \in \mathcal{H}_{PV+Bat} \cup \mathcal{H}_{Bat}$ decides when to charge or discharge the battery. The battery logic control for Off-Grid mode is given as follows:
\begin{align}
c_i(k) &= 
\begin{cases}
1,& \text{if } \bar{E}_i^{pv}(k) \geq E_i^{d_{2}}(k) \; , \\
0,& \text{otherwise} \; ,
\end{cases} \label{eq:Baseline_Charging} \\
d_i(k) &= 
\begin{cases}
1,& \text{if } E_i^{pv}(k) < E_i^{d_{2}}(k) \; , \\
0,& \text{otherwise} \; ,
\end{cases} \label{eq:Baseline_Discharging} \\
E_i^{d_{2}}(k) &= E_i^l(k) + u_i^{ac}(k) \cdot \bar{E}^{ac} \; .
\end{align}
Here $E_i^{d_{2}}(k)$ is the total house load demand. In this mode, the controller reflects the passive behavior of typical residential inverter-battery systems, where battery actions emerge from circuit-level interactions and a rated charge/discharge. The battery logic control for On-Grid mode is given as follows:
\begin{align}
c_i(k) &= 
\begin{cases}
1,& \text{if } E^{bat}(k)  < \bar{E}^{bat}(k) \; , \\
0,& \text{otherwise} \; ,
\end{cases} \label{eq:Baseline_Charging1} \\
d_i(k) &= 0 \label{eq:Baseline_Discharging1} \; .
\end{align}
Here $E_i^{d_{2}}(k)$ is the total house load demand. In this mode, the battery just tries to be fully charged without discharging.

\subsubsection{Load Control Logic}\label{subsubsec:LoadControlLogic}
The load logic controller for non-AC loads at house $i \in \mathcal{H}$ determines the status of the individual controllable loads and is given as follows:
\begin{align}
	u_i^{l,j}(k)= 
	\begin{cases}
		1,& \text{if } \bar{E}_i^{l,j}(k) > 0 \; ,\\
		0,& \text{if } \bar{E}_i^{l,j}(k) = 0 \; .
	\end{cases} \label{eq:LoadLogicControl}
\end{align}
This controller abstracts the physical on/off behavior of household devices, driven by occupants.

\subsection{Rule-Based Controller for Off-Grid Mode}\label{subsection:RuleBasedOffGrid}
\begin{algorithm}[hbt!]
\caption{Rule-Based Controller for Off-Grid Mode}
\label{alg:RuleBasedController}
\begin{algorithmic}[1]

\FORALL{$i \in \mathcal{H}$}
    \STATE Use the Solar PV model to compute $\bar{E}_i^{pv}(k)$ 
    \STATE Use the Baseline Controller (\ref{eq:Baseline_Charging})--(\ref{eq:LoadLogicControl}) to compute $u_i^{b}(k)$
    \STATE Compute $\delta_i^{ac}(k)$ using (\ref{eq:delta_ac})
\ENDFOR

\STATE Compute $E_{g}^{com}(k)$ and $E_{d}^{com}(k)$ using (\ref{eq:total_generation})--(\ref{eq:total_demand})
\STATE Compute energy mismatch $E_{mis}^{com}(k)$ using (\ref{eq:energy_mismatch})
\STATE Compute AC startup mismatch $P_{mis}^{ac}(k)$ using (\ref{eq:ac_startup_mismatch})

\IF{$E_{mis}^{com}(k) \geq 0$ \AND $P_{mis}^{ac}(k) \geq 0$}
    \FORALL{$i \in \mathcal{H}$}
        \STATE $u_i(k) \gets u_i^{b}(k)$ (use baseline controller commands)
    \ENDFOR
\ELSIF{$E_{mis}^{com}(k) \geq 0$ \AND $P_{mis}^{ac}(k) < 0$}
    \FORALL{$i \in \mathcal{H}$}
        \STATE $u_i^{ac}(k) \gets 0$ (turn off an appropriate portion of turning-on ACs)
        \STATE $u_i^{bat}(k) \gets u_i^{b,bat}(k)$ (use baseline controller commands)
        \STATE $u_i^{l}(k) \gets u_i^{b,l}(k)$ (use baseline controller commands)
    \ENDFOR
\ELSIF{$E_{mis}^{com}(k) < 0$}
    \FORALL{$i \in \mathcal{H}$}
        \IF{$\left| E_{mis}^{com}(k) \right| \leq \sum_{\mathcal{H}} u_i^{b,ac}(k) \cdot \bar{E}^{ac}$}
            \STATE $u_i^{ac}(k) \gets 0$ (turn off an appropriate portion of turning-on and turned-on ACs)
            \STATE $u_i^{bat}(k) \gets u_i^{b,bat}(k)$ (use baseline controller commands)
            \STATE $u_i^{l}(k) \gets u_i^{b,l}(k) $ (use baseline controller commands)
        \ELSIF{$\left| E_{mis}^{com}(k) \right| \leq \sum_{\mathcal{H}} u_i^{b,ac}(k) \cdot \bar{E}^{ac} + c_{i}^{b}(k) \cdot E_{i}^{bat,\bar{c}}$}
            \STATE $u_i^{ac}(k) \gets 0$ (turn off all turning-on and turned-on ACs)
            \STATE $u_i^{bat}(k) \gets 0$ (idle an appropriate portion of charging batteries)
            \STATE Compute $u_i^{ps,l}(k)$ using priority stack logic (\ref{eq:PriorityStackLogic})
            \STATE $u_i^{l}(k) \gets u_i^{ps,l}(k)$
        \ELSIF{$\left| E_{mis}^{com}(k) \right| \leq \sum_{\mathcal{H}} u_i^{b,ac}(k) \cdot \bar{E}^{ac} + c_{i}^{b}(k) \cdot E_{i}^{bat,\bar{c}}  + \bar{E}_i^{l}(k)$}
            \STATE $u_i^{ac}(k) \gets 0$ (turn off all turning-on and turned-on ACs)
            \STATE $u_i^{bat}(k) \gets [0,0]^{T}$ (idle all charging batteries)
            \STATE $u_i^{l}(k) \gets u_i^{ps,l}(k)$ (use priority stack controller)
        \ENDIF
    \ENDFOR
\ENDIF
\end{algorithmic}
\end{algorithm}

Algorithm~\ref{alg:RuleBasedController} presents a smarter alternative to the Baseline Controller for Off-Grid scenarios. While the Baseline Controller operates each device subsystem independently without accounting for community-level feasibility, the Rule-Based Controller improves upon this by using model information and mismatch computations to coordinate local decisions across the entire community. This controller inherits the logic of the baseline but overrides it when necessary to ensure feasibility. The priority order for curtailment is: turning-on ACs, turned-on ACs, charging batteries, and prioritized non-AC loads.

For each house $i \in \mathcal{H}$ it first computes its baseline control command $u_i^b(k) = [u_i^{b,ac}(k),u_i^{b,pv}(k),u_i^{b,bat}(k), u_i^{b,l}(k)]^{T}$, where $u_i^{b,bat}(k) = [c_i^b(k), d_i^b(k)]^{T}$ is the battery command, $u_i^{b,ac}(k)$ is the AC command, and $u_i^{b,l}(k)$ (all control notations in Algorithm~\ref{alg:RuleBasedController} without $b$ in the superscript are the realized control commands) is the vector of non-AC load commands. When curtailment of battery charging is required, the maximum charging dispatch at house $i$ is computed as follows:
\begin{align}
E_i^{bat,\bar{c}}(k) = \min \left\{ \frac{\bar{E}^{bat} - E_i^{bat}(k)}{\eta_{bat}^c}, \bar{E}^{bat,c} \right\}
\end{align}

If additional curtailment is still needed after disabling ACs and battery charging, the controller invokes a Priority Stack Controller to select the most essential loads. It turns on the prioritized loads so as to shed the remaining mismatch $E_{mis}^{l}$  as follows:
\begin{align}
    u_{i}^{l,j}(k) & =  
	\begin{cases}
		1, & \text{if } \sum_{\mathcal{H}} \sum_{m=1}^{j} \bar{E}_{i}^{l,m}(k) \leq | E_{mis}^{l}(k) - \\ 
        & \sum_{\mathcal{H}}  \sum_{m=j+1}^{8} \bar{E}_{i}^{l,m}(k) |   \; , \\
		0, & \text{otherwise} \; ,
	\end{cases} \label{eq:PriorityStackLogic}\\
    E_{mis}^{l}(k) &= |E_{mis}^{com}(k)| - \sum_{\mathcal{H}} \{ u_i^{b,ac}(k) \cdot \bar{E}^{ac} + c_{i}^{b}(k) \cdot E_{i}^{bat,\bar{c}} \} \; .
\end{align}
The resulting complete load control command is given by $u_l^{ps}(k) = [u_{i}^{l,1}(k), \dots, u_{|\mathcal{H}|}^{l,8\cdot|\mathcal{H}|}(k)]^{T}$.

\section{Case Study}\label{section:CaseStudy}

\begin{figure*}[htbp]
    \centering

    \begin{subfigure}[t]{0.48\textwidth}
        \centering
        \includegraphics[width=\linewidth]{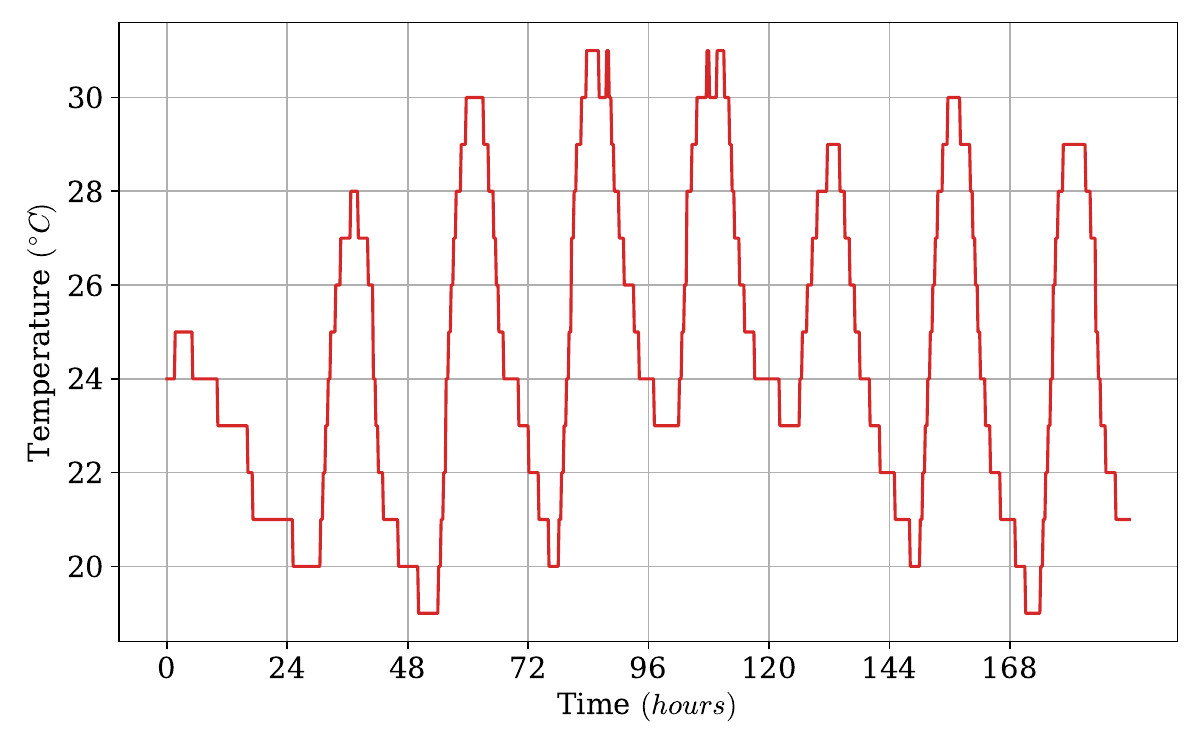}
        \caption{Temperature}
        \label{fig:Temperature}
    \end{subfigure}
    \hfill
    \begin{subfigure}[t]{0.48\textwidth}
        \centering
        \includegraphics[width=\linewidth]{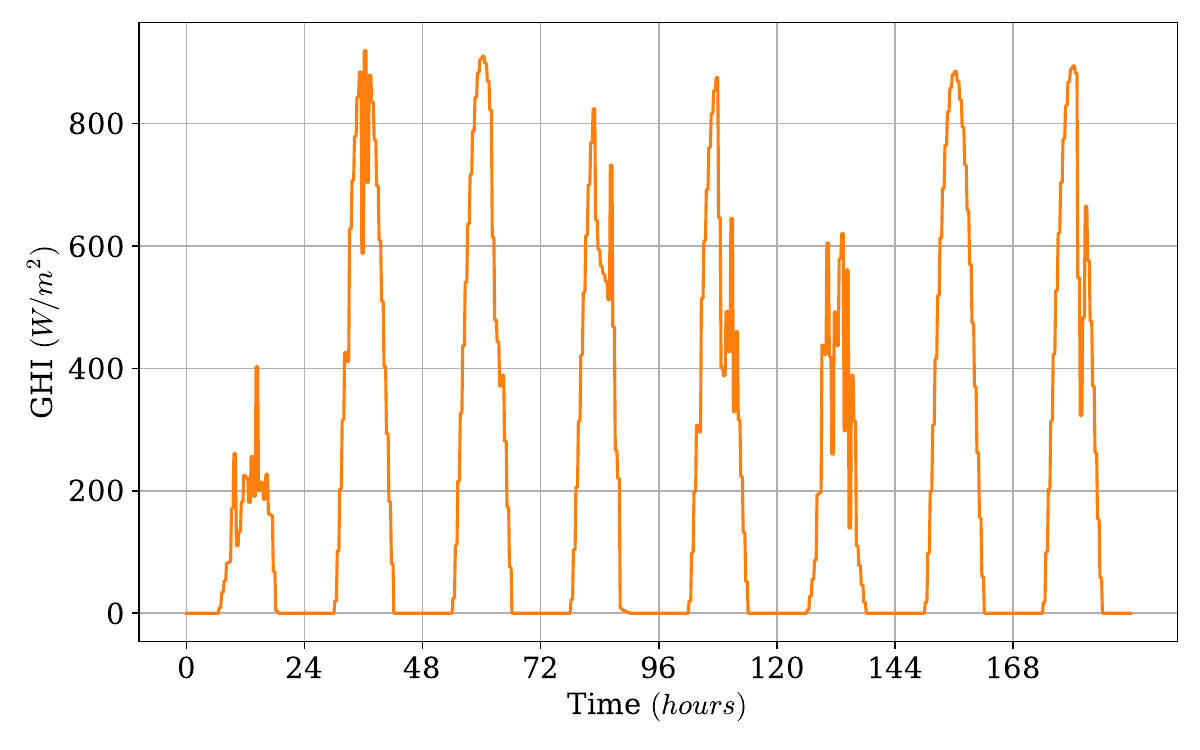}
        \caption{Global Horizontal Irradiance (GHI)}
        \label{fig:GHI}
    \end{subfigure}

    \vspace{0.8em} 

    \begin{subfigure}[t]{0.48\textwidth}
        \centering
        \includegraphics[width=\linewidth]{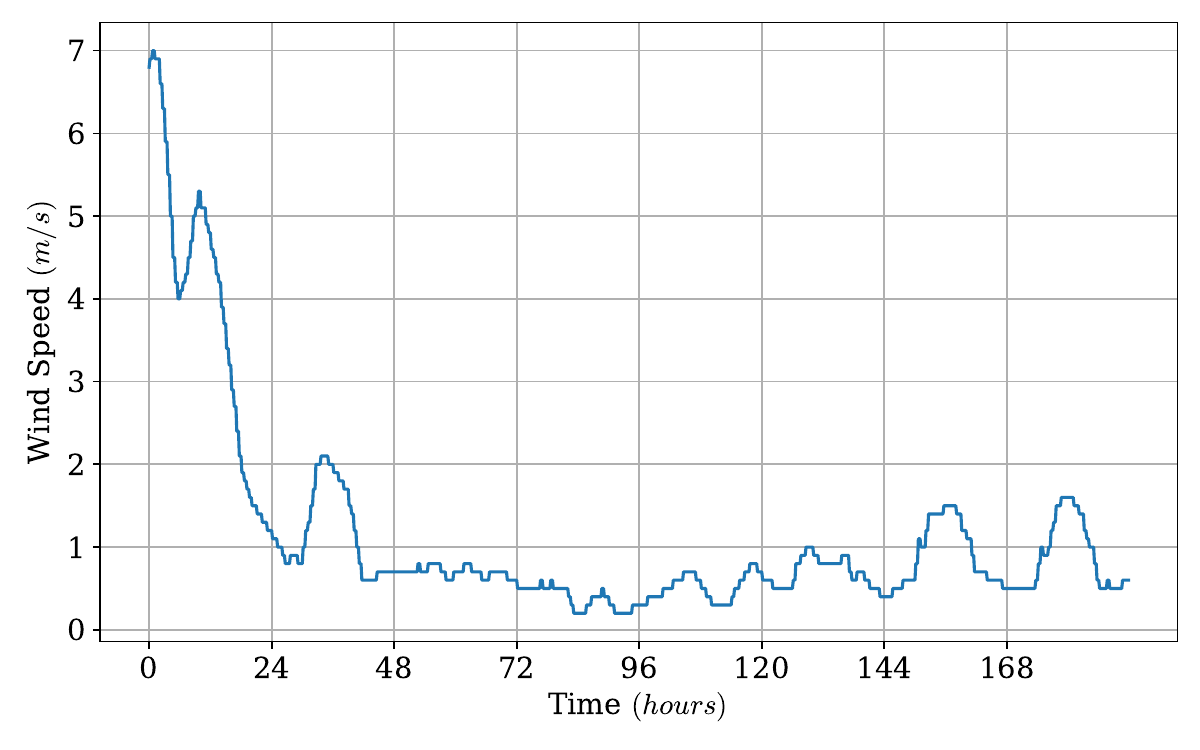}
        \caption{Wind Speed}
        \label{fig:Windspeed}
    \end{subfigure}
    \hfill
    \begin{subfigure}[t]{0.48\textwidth}
        \centering
        \includegraphics[width=\linewidth]{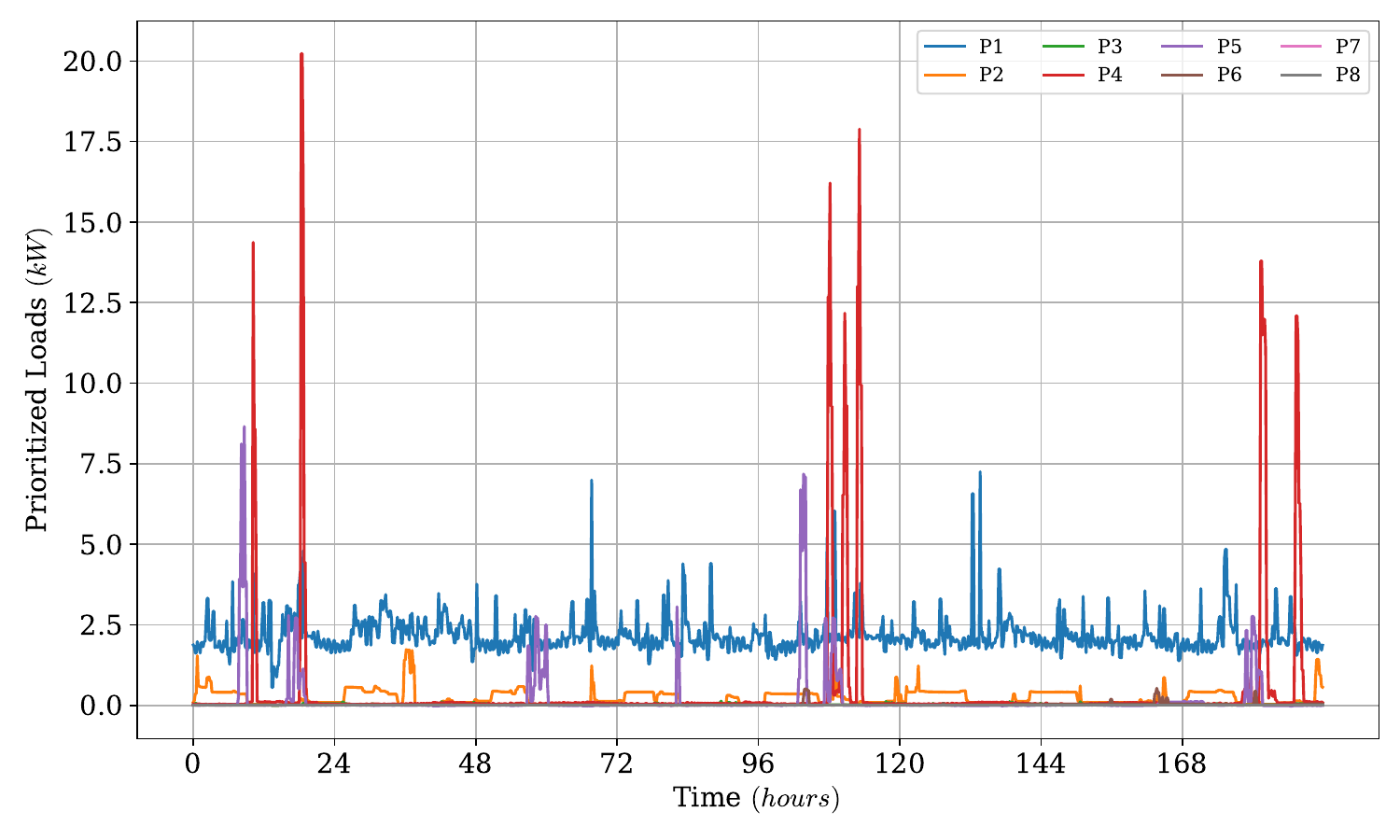}
        \caption{Load Demand}
        \label{fig:LoadDemand}
    \end{subfigure}

    \caption{Disturbance data used in the case study, including ambient temperature, solar irradiance (GHI), wind speed, and prioritized load demand profiles.}
    \label{fig:CaseStudy_Disturbances}
    \vspace{-0.4cm}
\end{figure*}

\subsection{Simulation Setup}\label{subsection:SimulationSetup}
We conduct closed-loop simulations of the Smart Community Simulator using each of the three controllers introduced in Section~\ref{section:DefaultControllers}. These simulations serve to highlight the capabilities of the simulator while enabling a comparative evaluation of controller performance across defined metrics. The simulation setup is described as follows:

\subsubsection{Simulation Cases}\label{subsec:SimulationCases}
We simulate five different community configurations. The first, labeled Community (1PV+Bat, 1Bat, 1PV, 1No-DER), consists of four houses—each with a distinct DER setup. The remaining four configurations are single-house cases: one house with PV and battery (1PV+Bat), one with only a battery (1Bat), one with only PV (1PV), and one with no DERs (1No-DER). Each configuration is evaluated under two grid connectivity modes: Off-Grid and On-Grid. For Off-Grid simulations, we consider two controller types—Baseline (BL) and Rule-Based (RB)—and two AC startup modes: With AC Startup Constraints (WACSC) and Without AC Startup Constraints (WOACSC). For the On-Grid mode, simulations are performed using the Baseline controller with the WOACSC startup configuration only. This results in a total of 25 distinct simulation cases: $5$ configurations $\times$ $(2 \times 2 ; \text{Off-Grid} + 1 ; \text{On-Grid})$. A Simulation Case refers to a unique combination of community configuration, controller, AC startup mode, and grid mode, whereas a Simulation Type refers to a unique combination of controller, startup mode, and grid mode, excluding the community configuration.

\subsubsection{Simulation Parameters}\label{subsec:SimulationParameters}
The PV system has a total capacity of $P_R^{pv} = 10.075 \; kW$ (Tesla SC325 modules, $0.325 \; kW$ each). The battery system (Tesla Powerwall) is modeled with $\bar{E}^{bat} = 13.5 \; kWh$, $\ubar{E}^{bat} = 0 \; kWh$, $\bar{E}^{bat,c} = 5 \cdot \Delta T_{s} \; kWh$, and $\bar{E}^{bat,d} = 7 \cdot \Delta T_{s} \; kWh$. The AC system has $P^{ac}_{R} = 3 \; kW$. Temperature bounds are $\ubar{T}^{ac} = 23  ^{\circ}C$ and $\bar{T}^{ac} = 25  ^{\circ}C$, and the heating/cooling mode switches at $\ubar{T}^{h} = 18  ^{\circ}C$ and $\bar{T}^{h} = 30  ^{\circ}C$. AC startup power is given by $\bar{P}^{ac}_{su} = (1 - \alpha_V) \cdot \alpha_I \cdot P^{ac}_{R} \; kW$, with $\alpha_V = 0.3$ and $\alpha_I = 5$, see~\cite{AC:LRA}. Thermal parameters follow~\cite{cui2019hybrid}.

\subsubsection{Simulation Period and Location}\label{subsec:TimeDuration}
The simulation spans a seven-day period from September 11 to September 17, 2017, with a time step of 10 minutes ($\Delta T_s = 10$ minutes or $\Delta T_s = 1/6$ hours). It models the behavior of a house located in Gainesville, FL, USA, during the passage of Hurricane Irma.

\subsubsection{Weather and Load Demand Data}\label{subsec:WeatherLoadData}
Weather data for the specified location and time period is obtained from the National Solar Radiation Database~\cite{NSRDBDatabase}, while non-thermal load demand data is sourced from the Pecan Street Dataport~\cite{PecanStreetDataport}, these for the disturbance data for the simulation see Fig.~\ref{fig:CaseStudy_Disturbances}.

\subsubsection{Computation}\label{subsec:Computation}
Simulations are conducted on a Windows desktop equipped with 16 GB RAM, a 2 GHz processor, and 8 CPU cores. The core functionality of the simulator and controllers is implemented in MATLAB, while the overall simulation is managed through \texttt{SmartCommunitySim}—a Python wrapper built on the Gymnasium environment, interfaced with MATLAB via the MATLAB Engine API.

\subsection{Results and Discussion}\label{subsection:ResultsDisscusion}

\begin{figure*}[!t]
    \centering

    \begin{subfigure}[t]{0.18\textwidth}
        \centering
        \includegraphics[width=\linewidth]{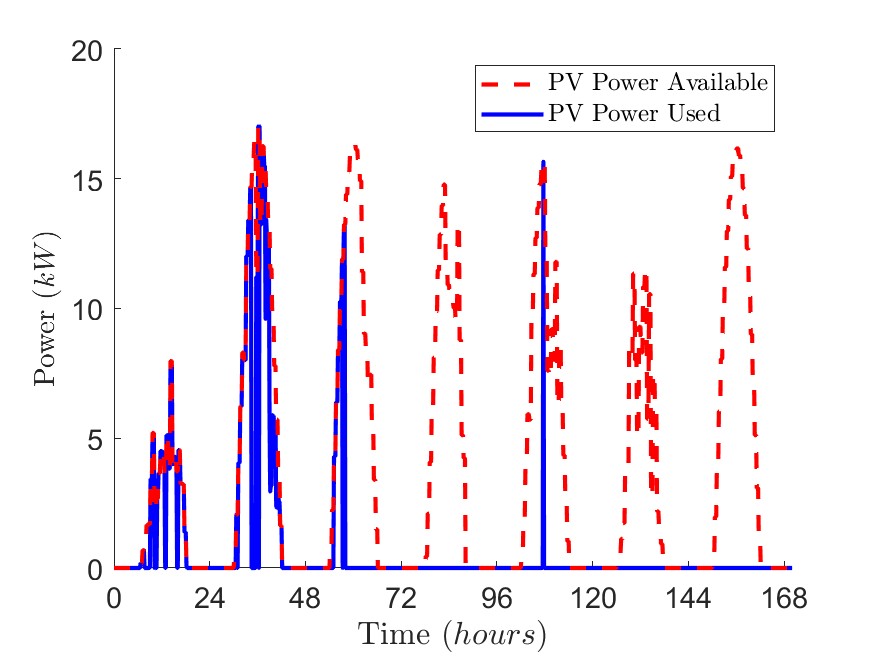}
        \caption{PV: Off-Grid Baseline (WACSC)}
        \label{fig:Com_PV_OG_BL_WA}
    \end{subfigure}
    \hfill
    \begin{subfigure}[t]{0.18\textwidth}
        \centering
        \includegraphics[width=\linewidth]{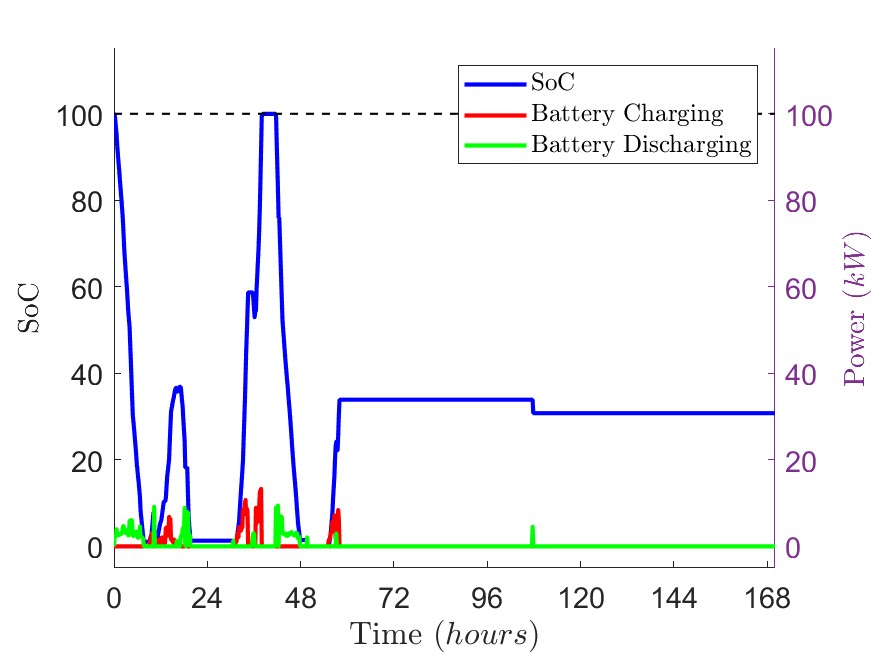}
        \caption{Battery: Off-Grid Baseline (WACSC)}
        \label{fig:Com_Bat_OG_BL_WA}
    \end{subfigure}
    \hfill
    \begin{subfigure}[t]{0.18\textwidth}
        \centering
        \includegraphics[width=\linewidth]{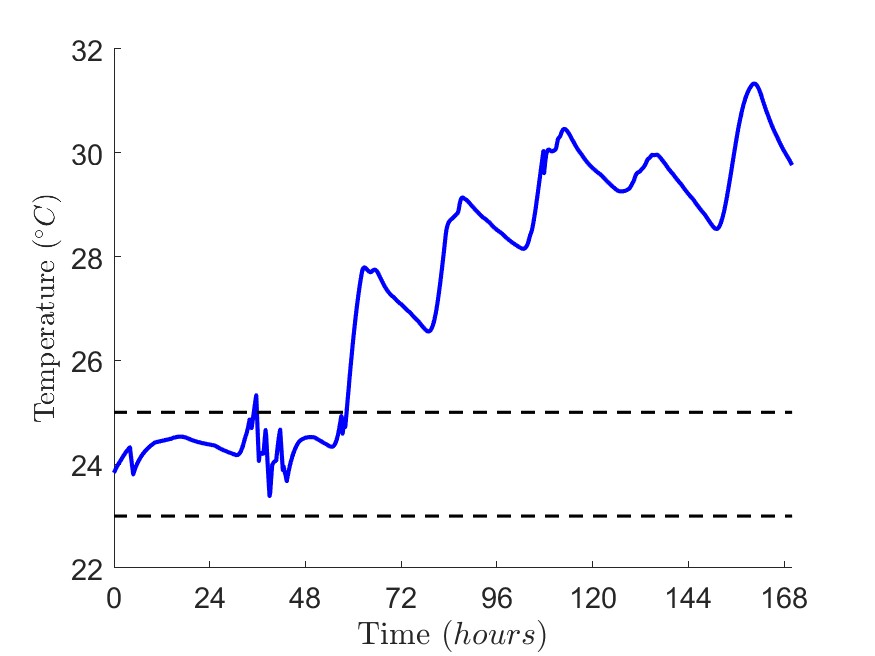}
        \caption{Temperature: Off-Grid Baseline (WACSC)}
        \label{fig:Com_T_OG_BL_WA}
    \end{subfigure}
    \hfill
    \begin{subfigure}[t]{0.18\textwidth}
        \centering
        \includegraphics[width=\linewidth]{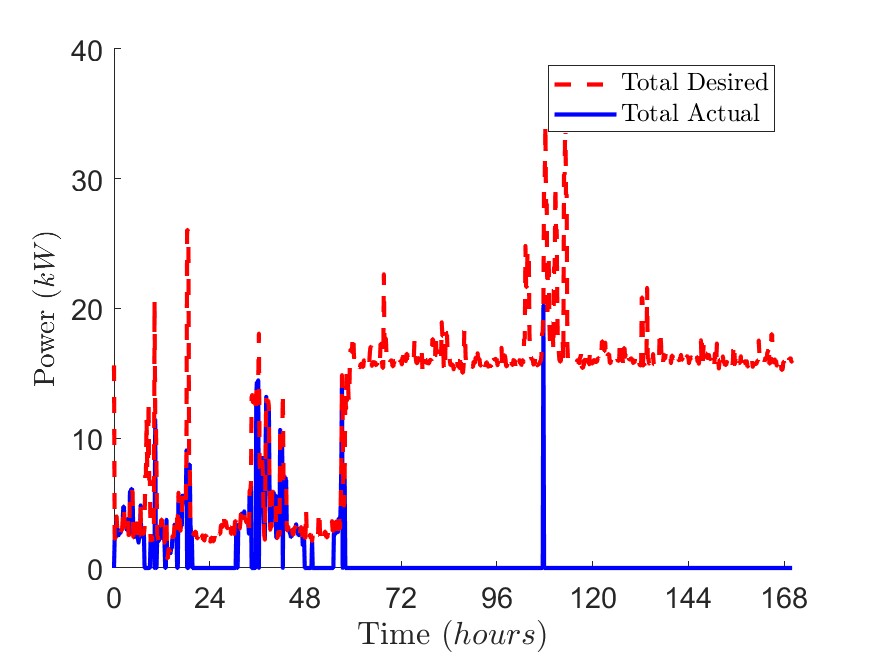}
        \caption{Load: Off-Grid Baseline (WACSC)}
        \label{fig:Com_L_OG_BL_WA}
    \end{subfigure}
    \hfill
    \begin{subfigure}[t]{0.18\textwidth}
        \centering
        \includegraphics[width=\linewidth]{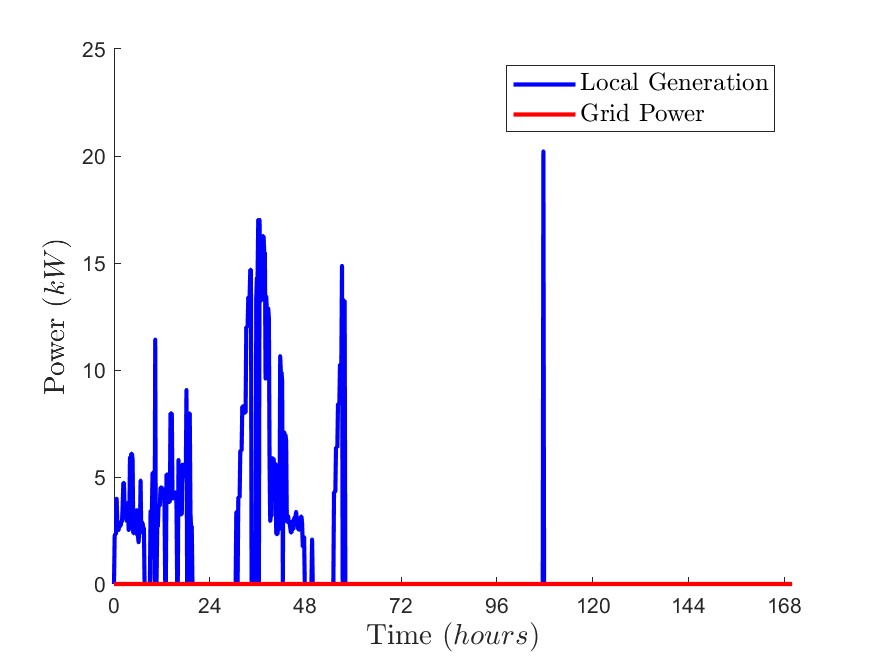}
        \caption{Generation: Off-Grid Baseline (WACSC)}
        \label{fig:Com_G_OG_BL_WA}
    \end{subfigure}

    \begin{subfigure}[t]{0.18\textwidth}
        \centering
        \includegraphics[width=\linewidth]{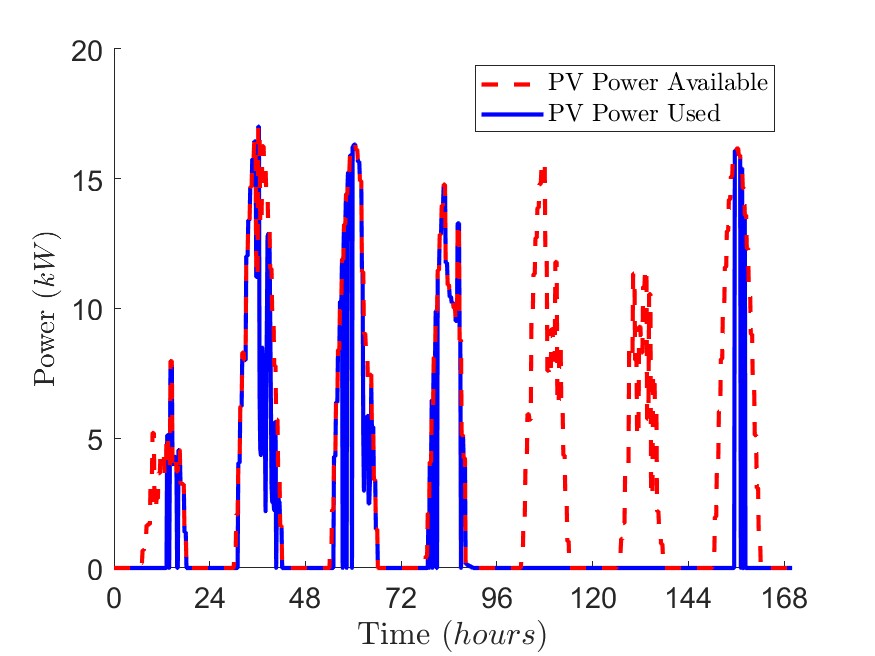}
        \caption{PV: Off-Grid Baseline (WOACSC)}
        \label{fig:Com_PV_OG_BL_WO}
    \end{subfigure}
    \hfill
    \begin{subfigure}[t]{0.18\textwidth}
        \centering
        \includegraphics[width=\linewidth]{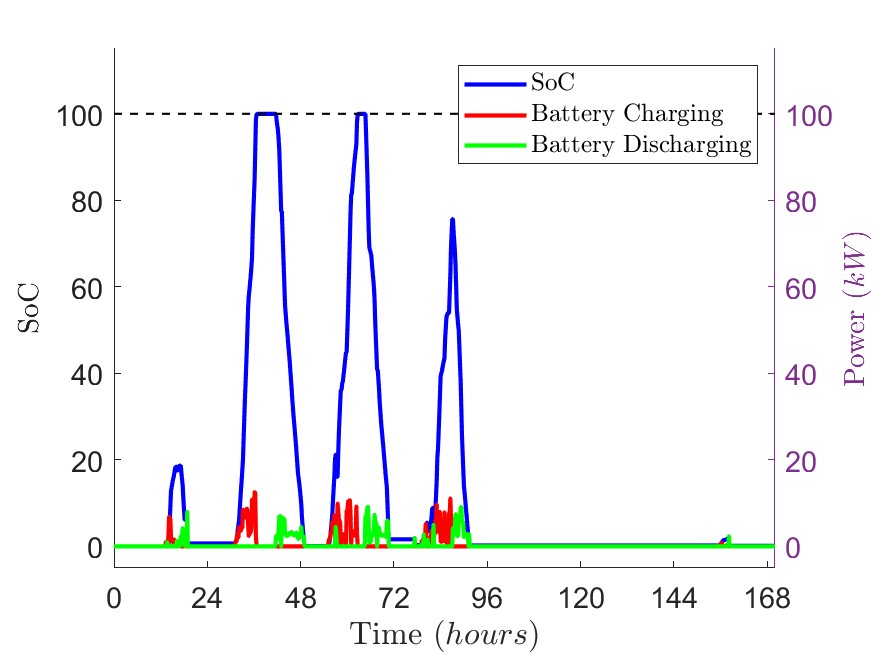}
        \caption{Battery: Off-Grid Baseline (WOACSC)}
        \label{fig:Com_Bat_OG_BL_WO}
    \end{subfigure}
    \hfill
    \begin{subfigure}[t]{0.18\textwidth}
        \centering
        \includegraphics[width=\linewidth]{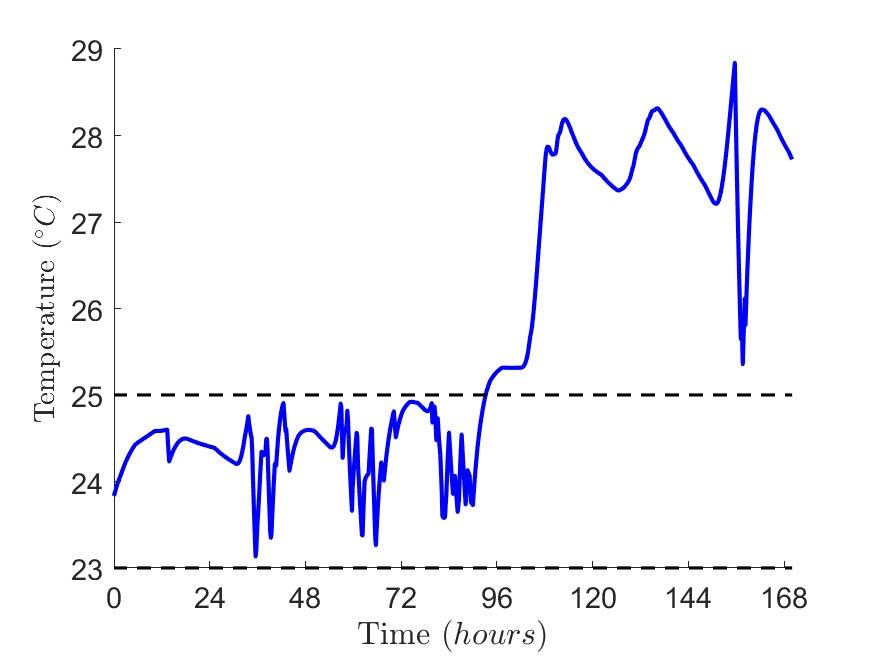}
        \caption{Temperature: Off-Grid Baseline (WOACSC)}
        \label{fig:Com_T_OG_BL_WO}
    \end{subfigure}
    \hfill
    \begin{subfigure}[t]{0.18\textwidth}
        \centering
        \includegraphics[width=\linewidth]{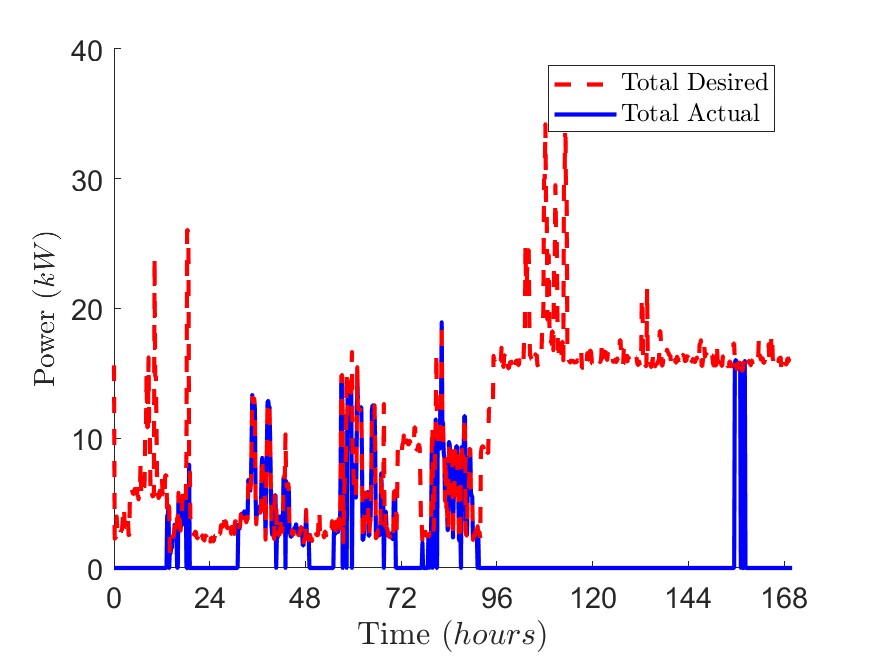}
        \caption{Load: Off-Grid Baseline (WOACSC)}
        \label{fig:Com_L_OG_BL_WO}
    \end{subfigure}
    \hfill
    \begin{subfigure}[t]{0.18\textwidth}
        \centering
        \includegraphics[width=\linewidth]{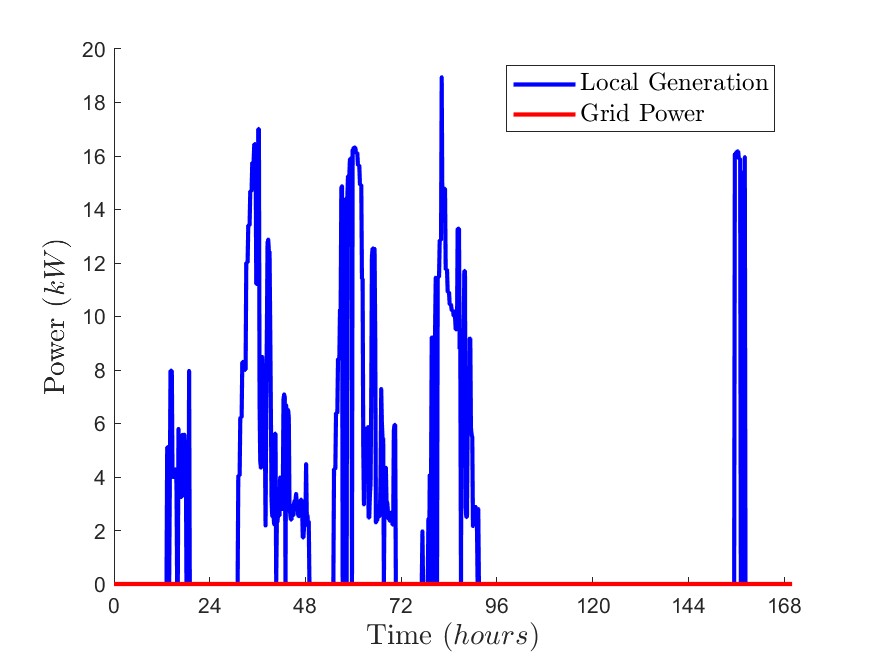}
        \caption{Generation: Off-Grid Baseline (WOACSC)}
        \label{fig:Com_G_OG_BL_WO}
    \end{subfigure}
    
    \vspace{0.5cm}
    \begin{subfigure}[t]{0.18\textwidth}
        \centering
        \includegraphics[width=\linewidth]{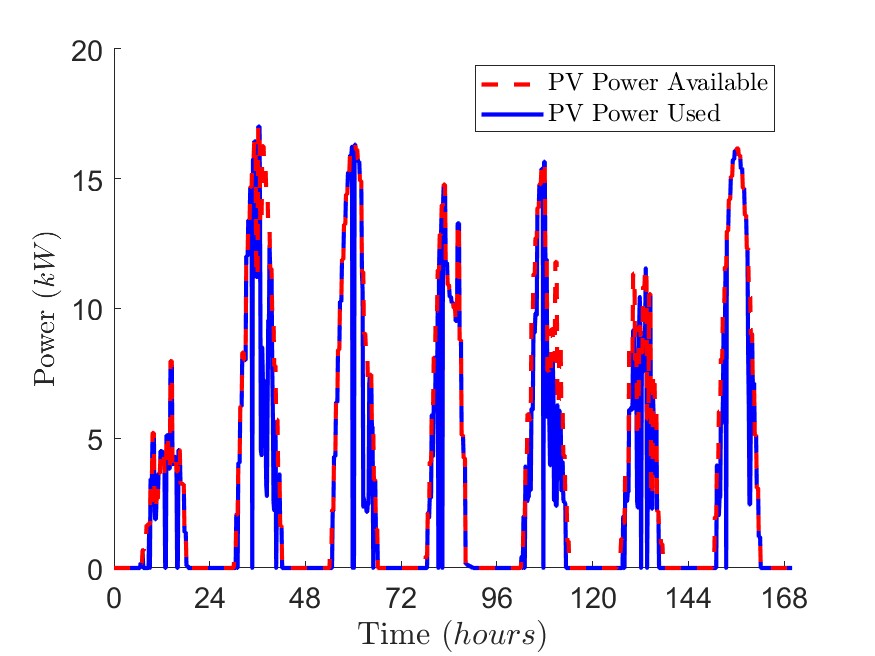}
        \caption{PV: Off-Grid Rule-Based (WACSC)}
        \label{fig:Com_PV_OG_RB_WA}
    \end{subfigure}
    \hfill
    \begin{subfigure}[t]{0.18\textwidth}
        \centering
        \includegraphics[width=\linewidth]{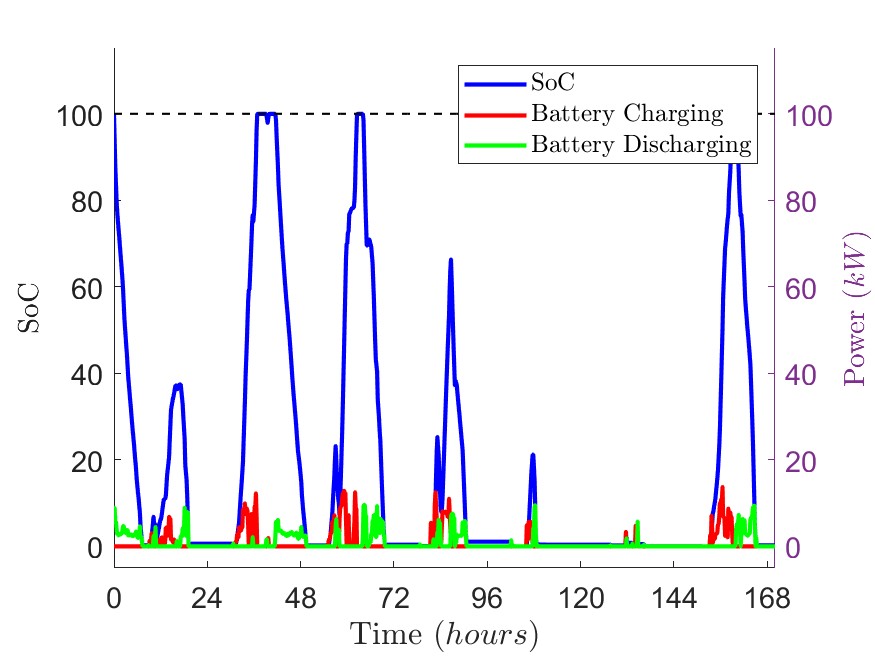}
        \caption{Battery: Off-Grid Rule-Based (WACSC)}
        \label{fig:Com_Bat_OG_RB_WA}
    \end{subfigure}
    \hfill
    \begin{subfigure}[t]{0.18\textwidth}
        \centering
        \includegraphics[width=\linewidth]{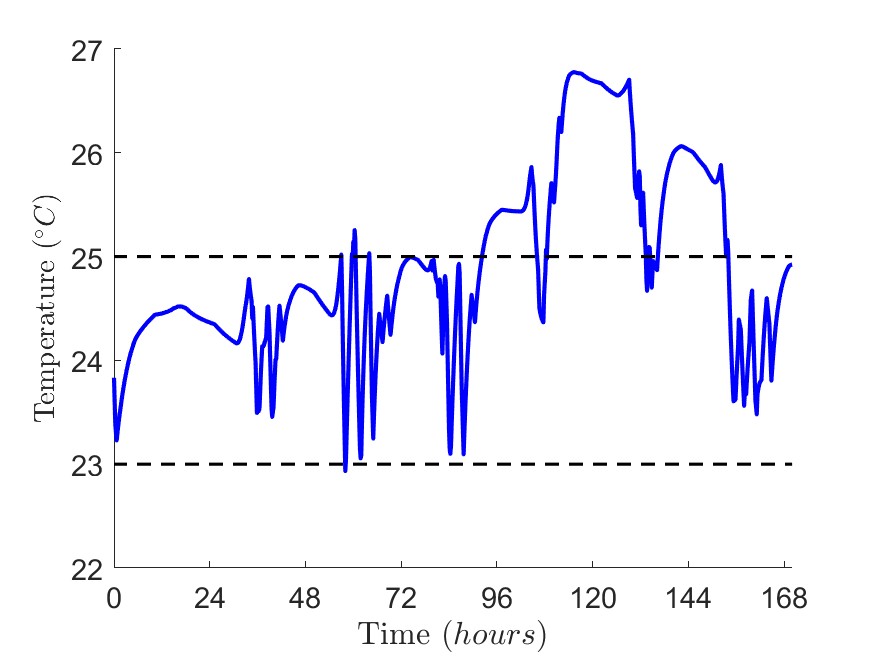}
        \caption{Temperature: Off-Grid Rule-Based (WACSC)}
        \label{fig:Com_T_OG_RB_WA}
    \end{subfigure}
    \hfill
    \begin{subfigure}[t]{0.18\textwidth}
        \centering
        \includegraphics[width=\linewidth]{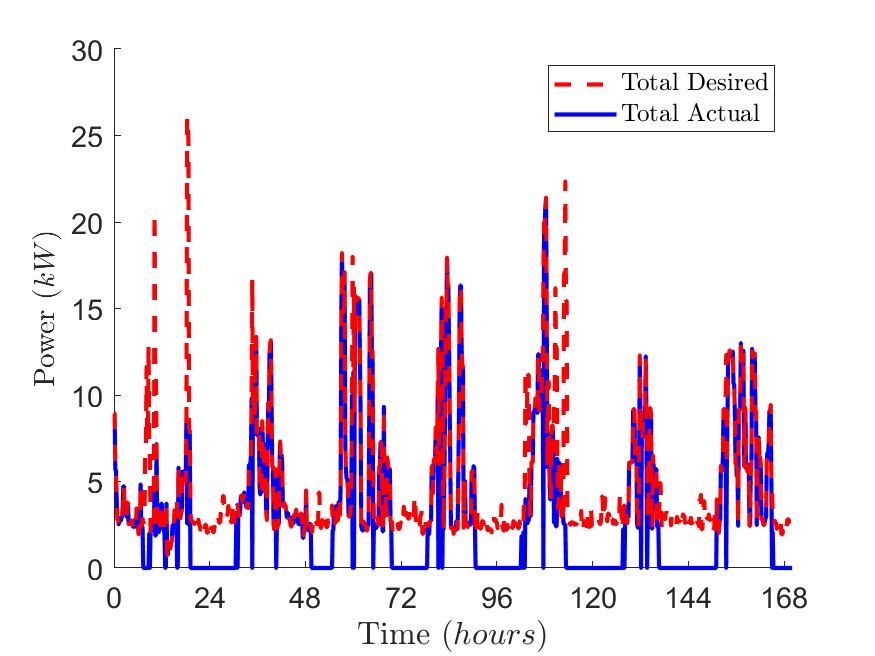}
        \caption{Load: Off-Grid Rule-Based (WACSC)}
        \label{fig:Com_L_OG_RB_WA}
    \end{subfigure}
    \hfill
    \begin{subfigure}[t]{0.18\textwidth}
        \centering
        \includegraphics[width=\linewidth]{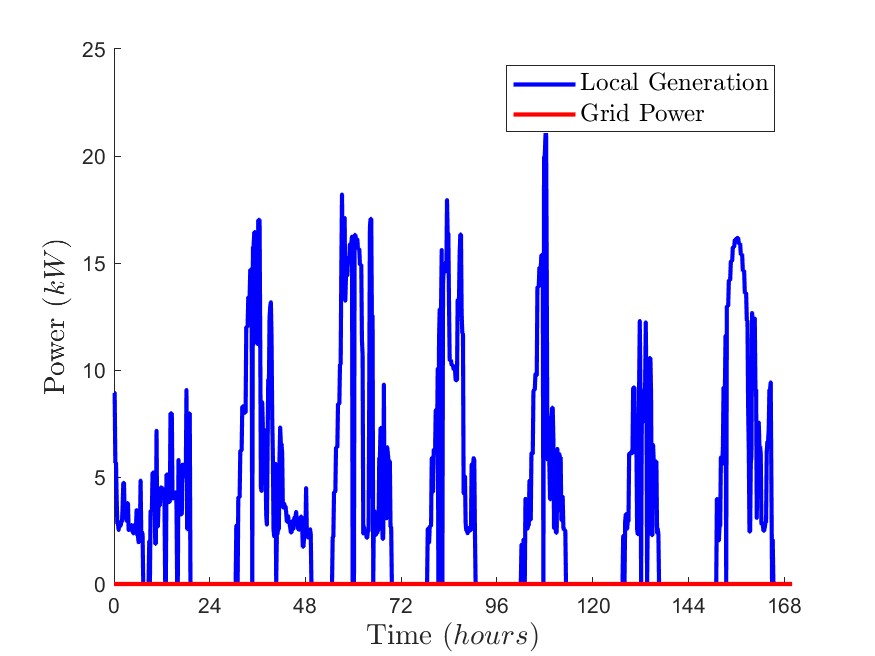}
        \caption{Generation: Off-Grid Rule-Based (WACSC)}
        \label{fig:Com_G_OG_RB_WA}
    \end{subfigure}

    \vspace{0.5cm}
    \begin{subfigure}[t]{0.18\textwidth}
        \centering
        \includegraphics[width=\linewidth]{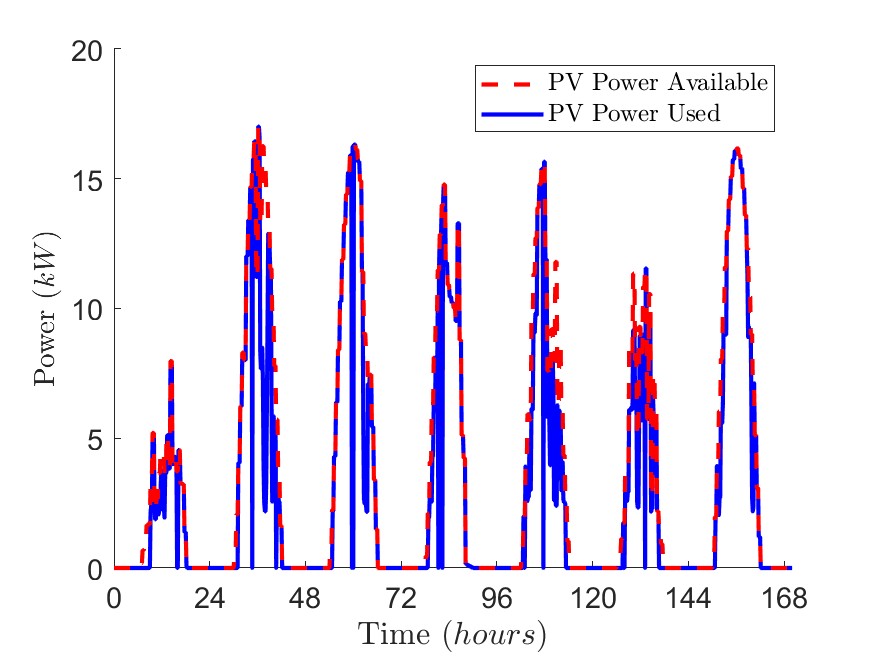}
        \caption{PV: Off-Grid Rule-Based (WOACSC)}
        \label{fig:Com_PV_OG_RB_WO}
    \end{subfigure}
    \hfill
    \begin{subfigure}[t]{0.18\textwidth}
        \centering
        \includegraphics[width=\linewidth]{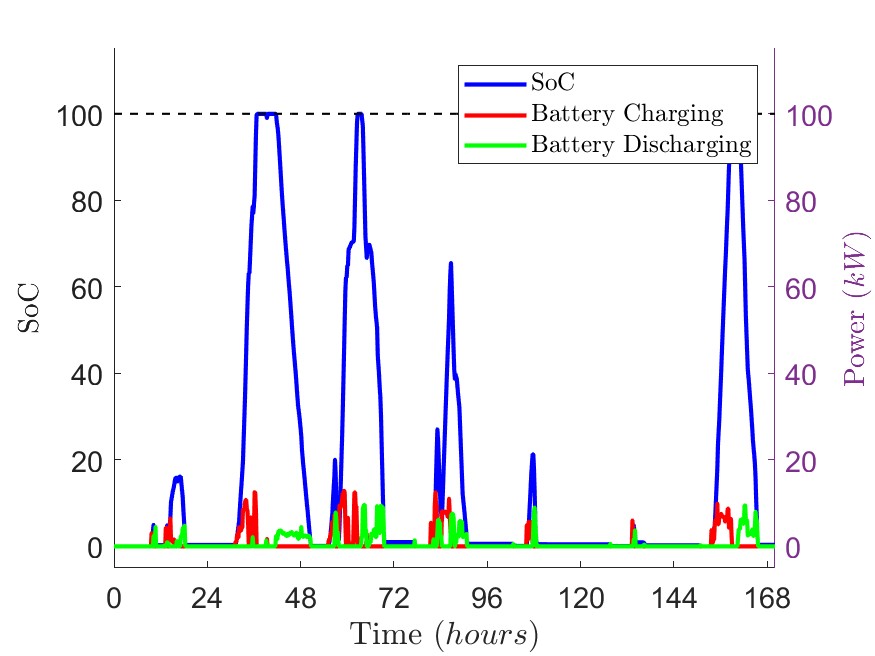}
        \caption{Battery: Off-Grid Rule-Based (WOACSC)}
        \label{fig:Com_Bat_OG_RB_WO}
    \end{subfigure}
    \hfill
    \begin{subfigure}[t]{0.18\textwidth}
        \centering
        \includegraphics[width=\linewidth]{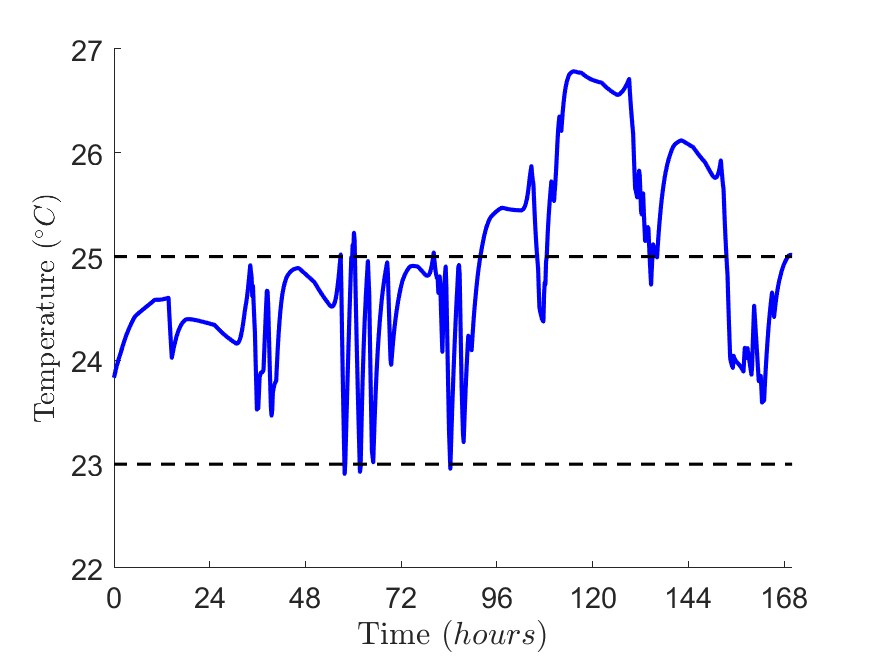}
        \caption{Temperature: Off-Grid Rule-Based (WOACSC)}
        \label{fig:Com_T_OG_RB_WO}
    \end{subfigure}
    \hfill
    \begin{subfigure}[t]{0.18\textwidth}
        \centering
        \includegraphics[width=\linewidth]{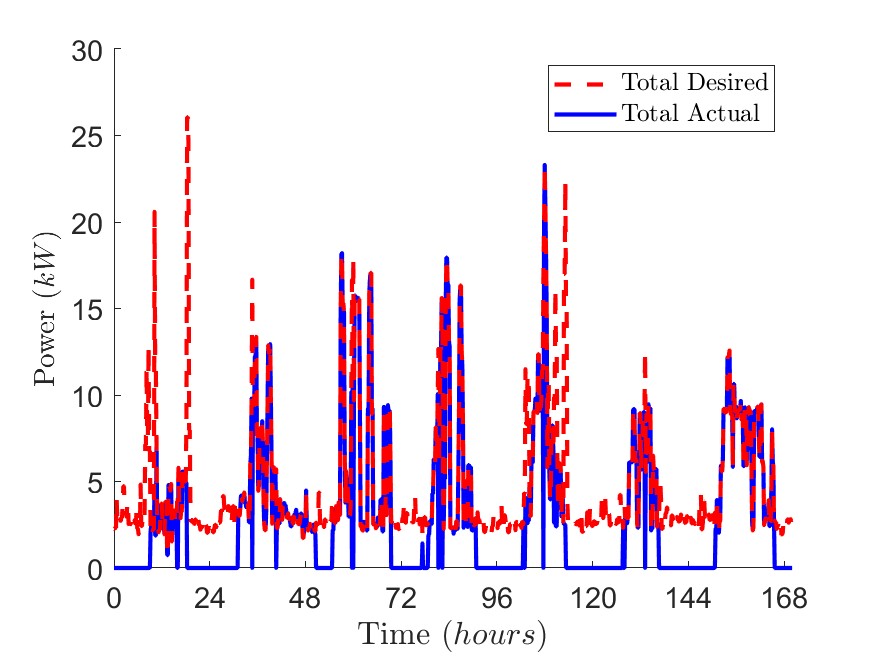}
        \caption{Load: Off-Grid Rule-Based (WOACSC)}
        \label{fig:Com_L_OG_RB_WO}
    \end{subfigure}
    \hfill
    \begin{subfigure}[t]{0.18\textwidth}
        \centering
        \includegraphics[width=\linewidth]{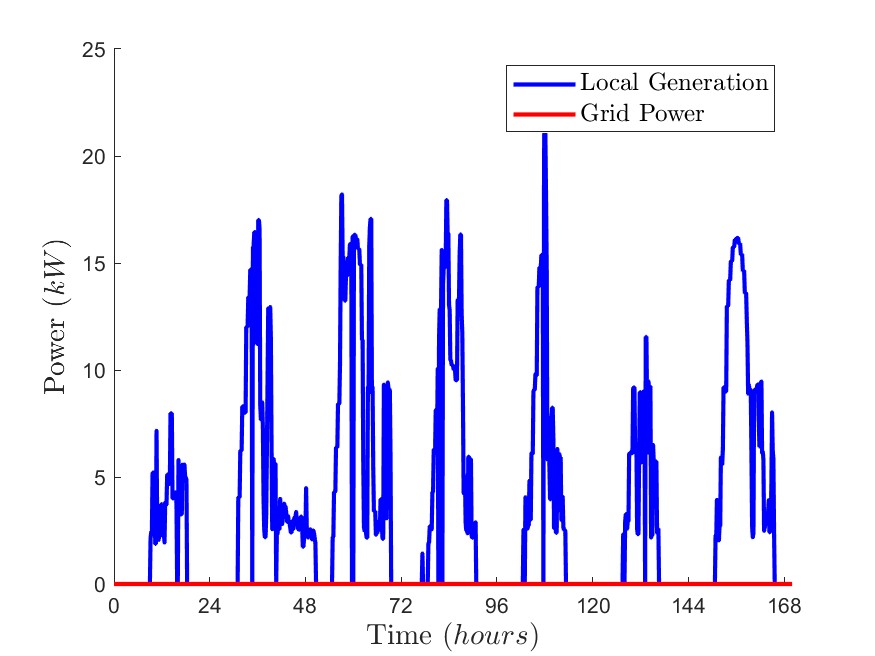}
        \caption{Generation: Off-Grid Rule-Based (WOACSC)}
        \label{fig:Com_G_OG_RB_WO}
    \end{subfigure}

    \vspace{0.5cm}
    \begin{subfigure}[t]{0.18\textwidth}
        \centering
        \includegraphics[width=\linewidth]{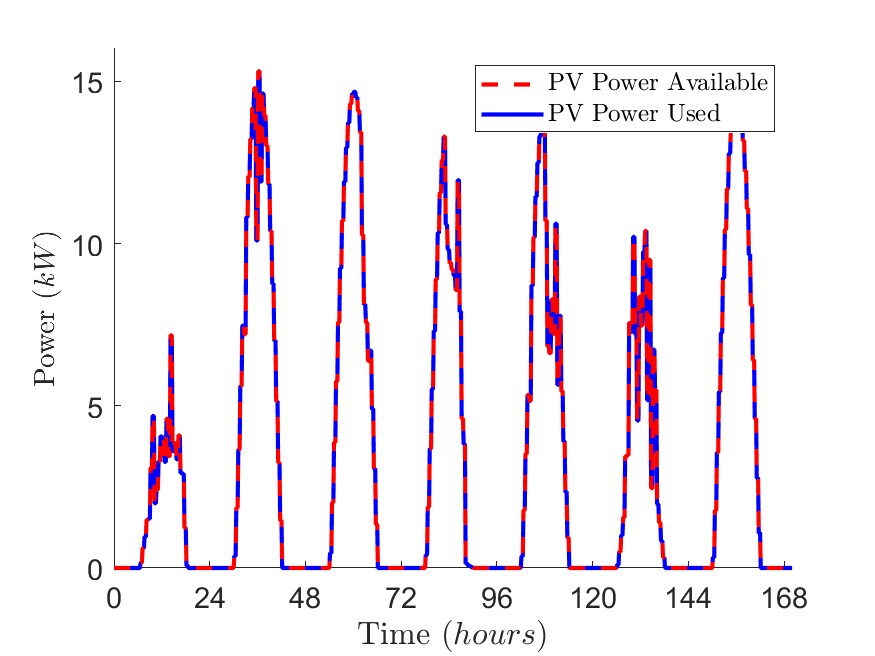}
        \caption{PV: On-Grid Baseline (WOACSC)}
        \label{fig:Com_PV_G_BL_WO}
    \end{subfigure}
    \hfill
    \begin{subfigure}[t]{0.18\textwidth}
        \centering
        \includegraphics[width=\linewidth]{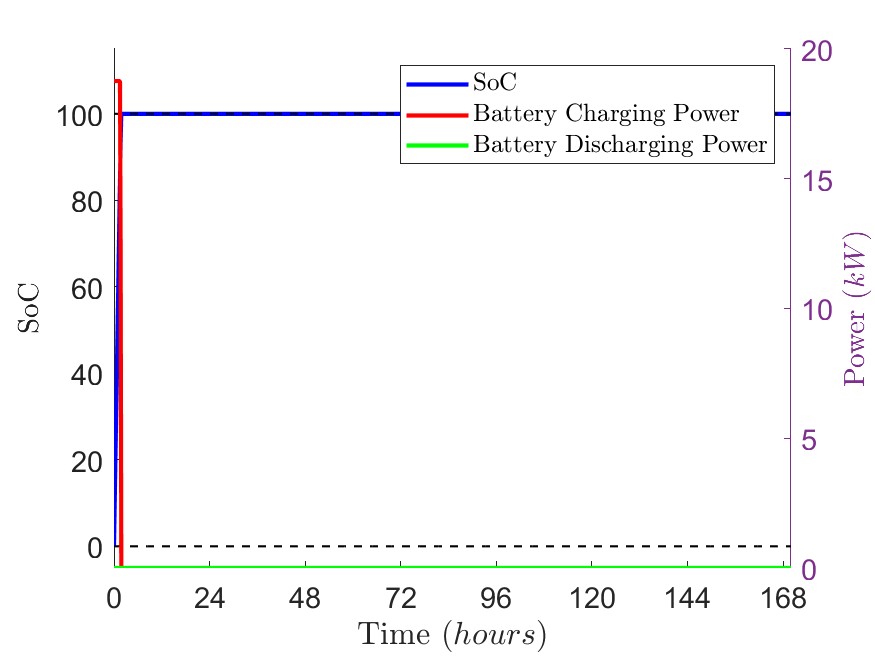}
        \caption{Battery: On-Grid Baseline (WOACSC)}
        \label{fig:Com_Bat_G_BL_WO}
    \end{subfigure}
    \hfill
    \begin{subfigure}[t]{0.18\textwidth}
        \centering
        \includegraphics[width=\linewidth]{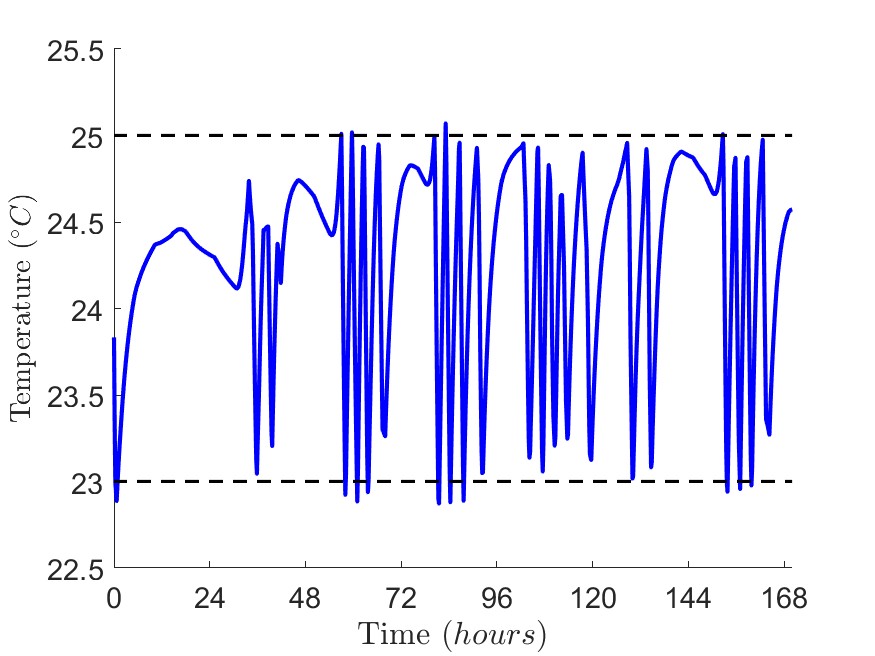}
        \caption{Temperature: On-Grid Baseline (WOACSC)}
        \label{fig:Com_T_G_BL_WO}
    \end{subfigure}
    \hfill
    \begin{subfigure}[t]{0.18\textwidth}
        \centering
        \includegraphics[width=\linewidth]{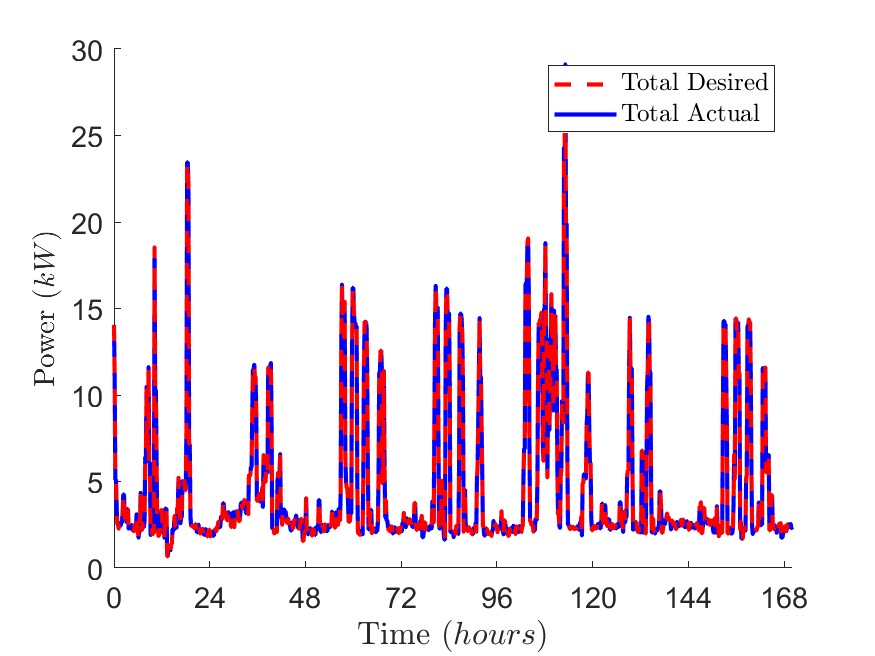}
        \caption{Load: On-Grid Baseline (WOACSC)}
        \label{fig:Com_L_G_BL_WO}
    \end{subfigure}
    \hfill
    \begin{subfigure}[t]{0.18\textwidth}
        \centering
        \includegraphics[width=\linewidth]{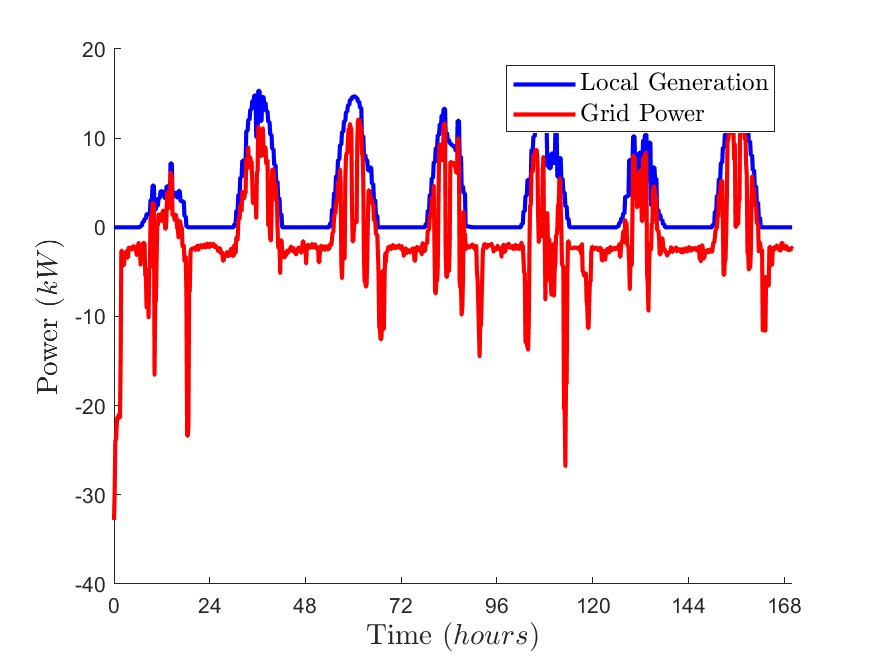}
        \caption{Generation: On-Grid Baseline (WOACSC)}
        \label{fig:Com_G_G_BL_WO}
    \end{subfigure}

    \caption{Community simulations across different Grid Modes, Controller Types, and AC Startup Modes.}
    \label{fig:Results}
\end{figure*}

\begin{figure*}[!t]
    \centering

    \begin{subfigure}[t]{0.48\textwidth}
        \centering
        \includegraphics[width=\linewidth]{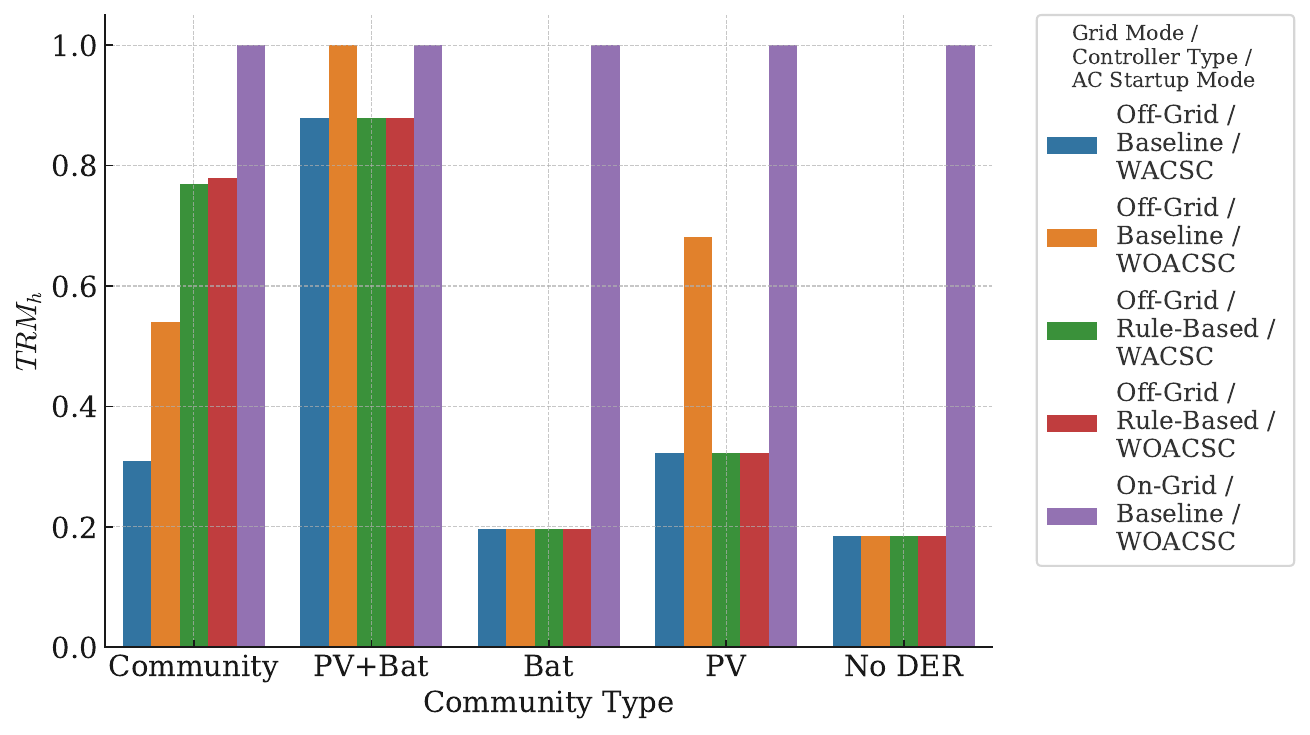}
        \caption{$TRM_{h}$}
        \label{fig:TRM_Metric}
    \end{subfigure}
    \hfill
    \begin{subfigure}[t]{0.48\textwidth}
        \centering
        \includegraphics[width=\linewidth]{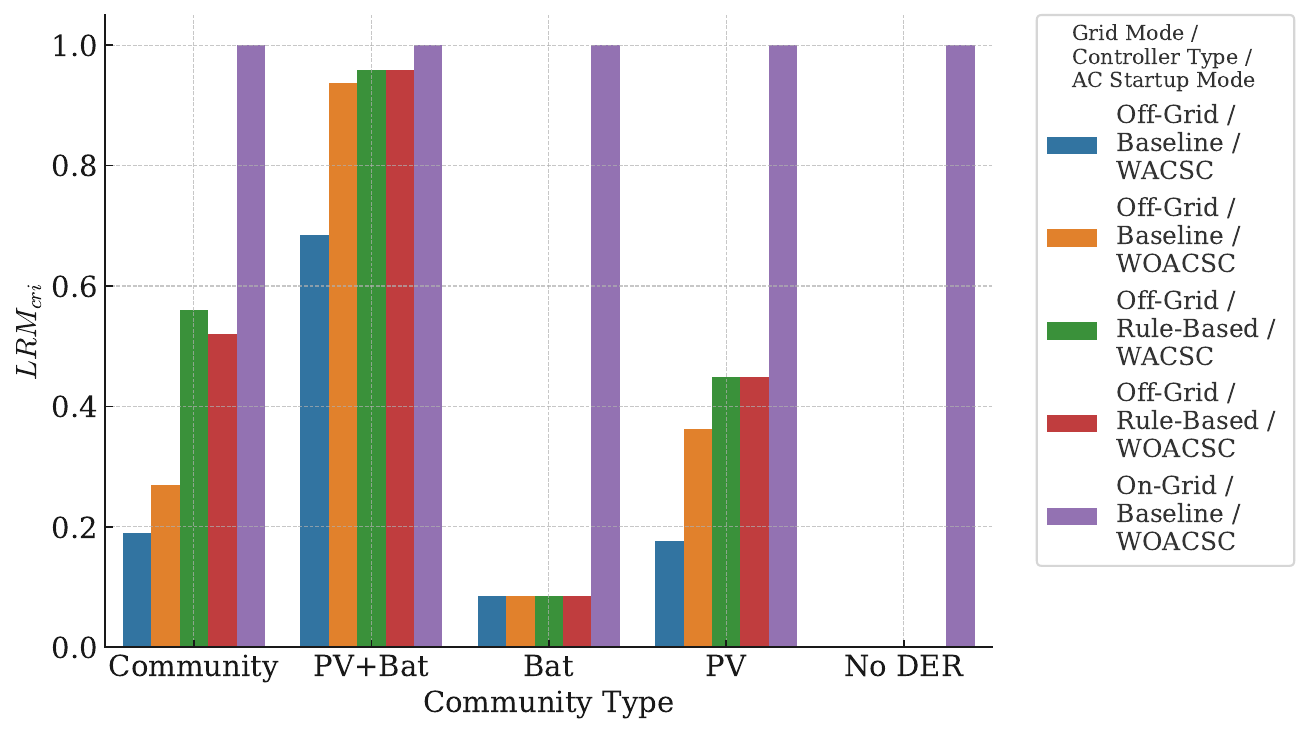}
        \caption{$LRM_{cri}$}
        \label{fig:LRM_C_Metric}
    \end{subfigure}

    \vspace{0.5em} 

    \begin{subfigure}[t]{0.48\textwidth}
        \centering
        \includegraphics[width=\linewidth]{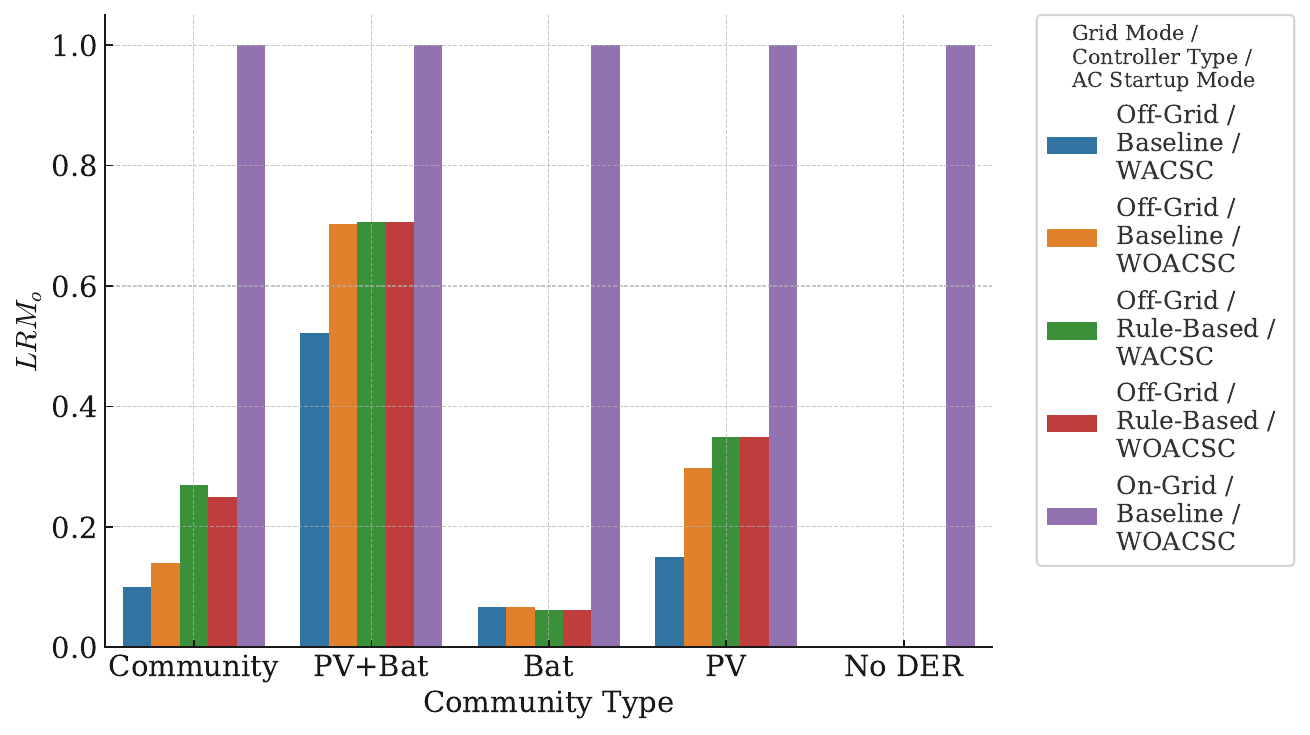}
        \caption{$LRM_{o}$}
        \label{fig:LRM_O_Metric}
    \end{subfigure}
    \hfill
    \begin{subfigure}[t]{0.48\textwidth}
        \centering
        \includegraphics[width=\linewidth]{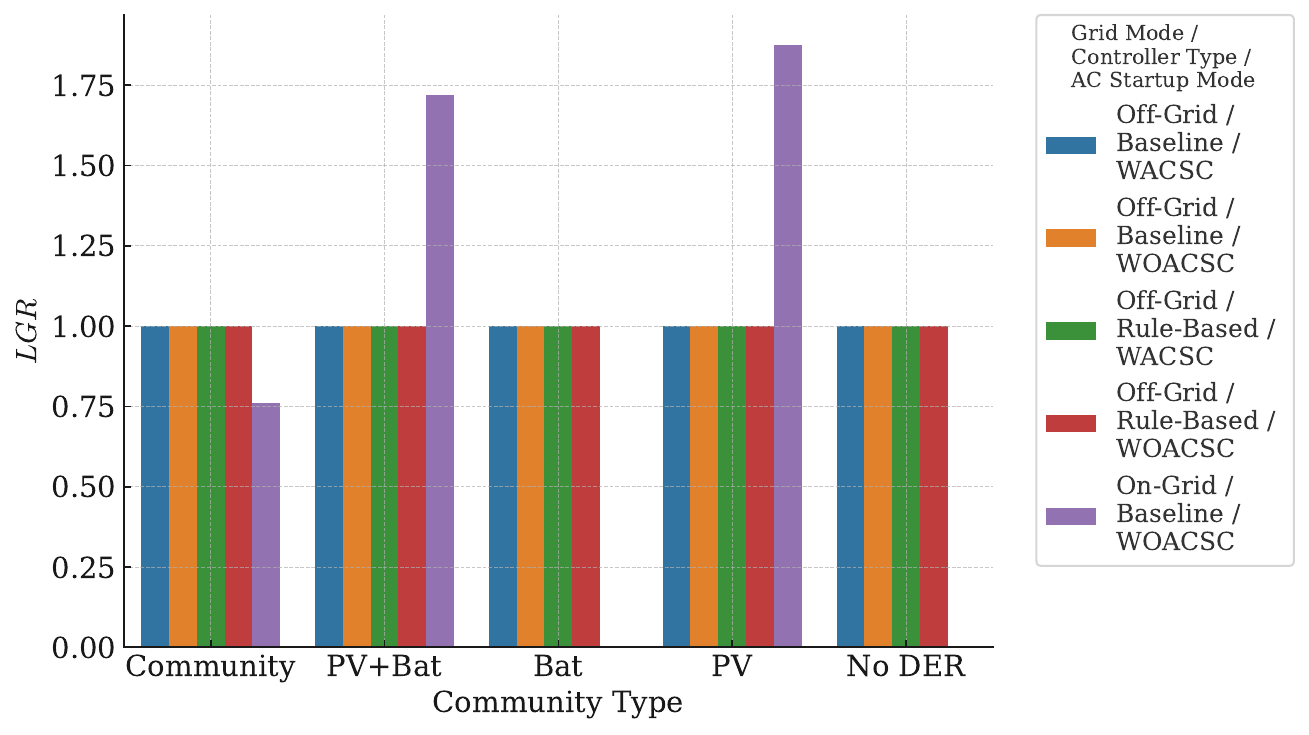}
        \caption{$LGR$}
        \label{fig:LGR_Metric}
    \end{subfigure}

    \caption{Comparison of performance metrics for all community types across all grid modes, controller types, and AC startup modes.}
    \label{fig:CaseStudy_Metrics}
    \vspace{-0.4cm}
\end{figure*}

\subsubsection{Performance Metrics}\label{subsubsection:PerformanceMetrics}
We evaluate the performance of each simulation using four key metrics that capture different aspects of system resiliency and energy sufficiency. These metrics are computed as community-level averages, aggregating values across all houses in the system. Specifically: (i) the Critical Load Resiliency Metric ($LRM_{cri}$) measures the fraction of critical loads (priority level $P1$) served over the simulation horizon; (ii) the Other Loads Resiliency Metric ($LRM_{o}$) quantifies the proportion of other loads served, including critical loads; (iii) the Thermal Resiliency Metric ($TRM_{h}$) captures average thermal comfort violations based on indoor temperature deviations from the temperature limits (iv) the Local Generation Ratio ($LGR$) measures the share of total community demand met using local DERs (PV and battery), without relying on grid imports. The mathematical definitions of these metrics are provided below:
\begin{align}
&LRM_{cri} = 
\frac{ \sum\limits_{k=1}^{T_{sim}} \left\{ \sum\limits_{i \in \mathcal{H}} E_{i}^{l,1}(k) \right\} }
     { \sum\limits_{k=1}^{T_{sim}} \left\{ \sum\limits_{i \in \mathcal{H}} \bar{E}_{i}^{l,1}(k) \right\} },
LRM_{o} = 
\frac{ \sum\limits_{k=1}^{T_{sim}} \left\{ \sum\limits_{i \in \mathcal{H}} E_{i}^{l}(k) \right\} }
     { \sum\limits_{k=1}^{T_{sim}} \left\{ \sum\limits_{i \in \mathcal{H}} \bar{E}_{i}^{l}(k) \right\} }, 
\\[10pt]
&TRM_{h} = 
\frac{1}{|\mathcal{H}| T_{sim}} \sum_{i \in \mathcal{H}} \sum_{k=1}^{T_{sim}} 
\left\{
\begin{aligned}
&\left[ \bar{T}^{h} - T_{i}^{h}(k) \right]_+ \\
&+ \left[ T_{i}^{h}(k) - \ubar{T}^{h} \right]_-
\end{aligned}
\right\}, \\
&LGR = 
\frac{ \sum\limits_{k=1}^{T_{sim}} E_{g}^{com}(k) }
     { \sum\limits_{k=1}^{T_{sim}} E_{d}^{com}(k) }.
\end{align}

Here, $T_{sim}$ is the total number of simulation time steps. 

\subsubsection{Comparison of all Simulation Types for Community}\label{subsubsection:ComparisonCommunity}
We compare the performance of the Community configuration (1PV+Bat, 1Bat, 1PV, 1No-DER) across all simulation types, defined by combinations of grid mode (Off-Grid or On-Grid), controller type (Baseline or Rule-Based), and AC startup constraints (WACSC or WOACSC). Fig.~\ref{fig:Results} (plots are created through the internal visualization capability of the simulator package) presents system behaviors for three representative cases: Off-Grid Baseline (WACSC), Off-Grid Rule-Based (WACSC), and On-Grid Baseline (WOACSC). Each row of subplots captures trends in PV availability and utilization, battery dispatch (charging/discharging and SoC), indoor temperature regulation, total load consumption versus served, and community-level generation-load balance. Table~\ref{tab:Results} reports thermal resiliency ($TRM_h$), critical load resiliency ($LRM_{cri}$), other loads resiliency ($LRM_o$), and local generation ratio ($LGR$).

In the Off-Grid Baseline (WACSC) scenario (Fig.~\ref{fig:Com_PV_OG_BL_WA}-Fig.~\ref{fig:Com_G_OG_BL_WA}), the PV is not fully utilized due to curtailment during midday peaks, and battery usage remains minimal after early depletion (Fig.~\ref{fig:Com_PV_OG_BL_WA}-Fig.~\ref{fig:Com_Bat_OG_BL_WA}). The indoor temperature (Fig.~\ref{fig:Com_T_OG_BL_WA}) continuously drifts above comfort thresholds due to insufficient HVAC operation, leading to a low $TRM_h = 0.31$. The load plot (Fig.~\ref{fig:Com_L_OG_BL_WA}) shows significant deviation between desired and actual served loads, reflected in $LRM_{cri} = 0.19$ and $LRM_o = 0.10$. By contrast, in the Off-Grid Rule-Based case (Fig.~\ref{fig:Com_PV_OG_RB_WA}-Fig.~\ref{fig:Com_G_OG_RB_WA}), the controller more efficiently dispatches PV and battery resources, maintains tighter temperature control (Fig.~\ref{fig:Com_PV_OG_RB_WA}-Fig.~\ref{fig:Com_Bat_OG_RB_WA}), and serves more loads (Fig.~\ref{fig:Com_L_OG_RB_WA}). This results in improved performance: $TRM_h = 0.77$, $LRM_{cri} = 0.56$, and $LRM_o = 0.27$.

In the On-Grid Baseline (WOACSC) case (Fig.~\ref{fig:Com_PV_G_BL_WO}-Fig.~\ref{fig:Com_G_G_BL_WO}), grid access guarantees nearly perfect load and temperature regulation, with HVAC systems cycling responsively (Fig.~\ref{fig:Com_T_G_BL_WO}) and no observable gaps between desired and served load (Fig.~\ref{fig:Com_L_G_BL_WO}). All resiliency metrics reach unity ($TRM_h = LRM_{cri} = LRM_o = 1$), but the $LGR = 0.76$ confirms that a significant portion of demand is met through grid imports. The generation-load balance plot (Fig.~\ref{fig:Com_G_G_BL_WO}) also clearly shows the dominance of red (grid power) compared to local generation (blue), further reinforcing this observation.

These results illustrate the critical impact of controller design and grid connectivity on community performance. While the Rule-Based controller improves Off-Grid operation through better prioritization and load coordination, its effectiveness is still limited by the physical constraints of local generation. The On-Grid scenario demonstrates superior performance in terms of all loads being serviced all the time due to the presence of the grid. Overall, this comparison shows that the Smart Community Simulator is a flexible and extensible framework for evaluating control strategies, house-DER configurations in different grid modes.  
\begin{table}[ht]
    \centering
    \caption{Performance Metrics for Community}
    \label{tab:Results}
    \begin{tabular}{|l|c|c|c|c|}
        \hline
        \textbf{Simulation Type} & \textbf{$TRM_{h}$} & \textbf{$LRM_{cri}$} & \textbf{$LRM_o$} & \textbf{$LGR$}  \\ \hline
        Off-Grid BL (WACSC) & 0.31 & 0.19 & 0.1 & 1 \\ \hline
        Off-Grid BL (WOACSC) & 0.54 & 0.27 & 0.14 & 1 \\ \hline
        Off-Grid RB (WACSC) & 0.77 & 0.56 & 0.27 & 1  \\ \hline
        Off-Grid RB (WOACSC) & 0.78 & 0.52 & 0.25 & 1  \\ \hline
        On-Grid BL (WOACSC) & 1 & 1 & 1 & 0.76 \\ \hline
    \end{tabular}
\end{table}

\subsubsection{Comparison of Performance Metrics for all Community Types across all Simulation Types}\label{subsubsection:ComparisonPerformanceMetrics}
We present a comparative analysis of all community types—Community, PV+Bat, PV, Bat, and No DER—under different combinations of grid mode, controller type, and AC startup mode. The performance is evaluated using four key metrics: thermal resiliency ($TRM_h$), critical load resiliency ($LRM_{cri}$), other loads resiliency ($LRM_o$), and local generation ratio ($LGR$), as shown in Fig.~\ref{fig:CaseStudy_Metrics}. Each community type is analyzed across all five simulation types: Off-Grid with Baseline/WACSC, Off-Grid with Baseline/WOACSC, Off-Grid with Rule-Based/WACSC, Off-Grid with Rule-Based/WOACSC, and On-Grid with Baseline/WOACSC.

As expected, the On-Grid Baseline (WOACSC) case yields perfect resiliency across all community types ($TRM_h = LRM_{cri} = LRM_o = 1$) due to unrestricted power availability. However, the local generation ratio ($LGR$) under this mode varies with DER penetration. For example, PV-only and PV+Bat configurations exhibit high $LGR$ values ($1.87$ and $1.72$ respectively), indicating significant self-consumption and grid export. In contrast, configurations like Bat-only or No DER show $LGR = 0$, confirming complete reliance on the grid. Under Off-Grid modes, the Rule-Based controller consistently outperforms the Baseline controller, particularly in thermal comfort ($TRM_h$) and load service metrics. For instance, PV+Bat achieves $TRM_h \approx 0.88$ under all Off-Grid simulation types, and $LRM_{cri}$ increases from $0.68$ (Baseline/WACSC) to $0.96$ (Rule-Based/WACSC or WOACSC). The Community configuration also benefits significantly from intelligent control, with $TRM_h$ improving from $0.31$ (Baseline/WACSC) to $0.78$ (Rule-Based/WOACSC), and $LRM_{cri}$ from $0.19$ to $0.56$.

These results show that increasing DER diversity and deploying intelligent control substantially improve performance in Off-Grid modes. The PV+Bat and Community configurations under Rule-Based control approach On-Grid-level resiliency, confirming the effectiveness of load prioritization and AC startup-aware dispatch. On the other hand, configurations with limited flexibility—like Bat-only or No DER—show poor resiliency in Off-Grid settings regardless of control type. This illustrates the importance of both DER heterogeneity and control sophistication for enhancing resilience and minimizing dependence on the grid. Finally, the built-in performance metrics provide a structured and quantitative basis for assessing simulation results and comparing control strategies.

\subsubsection{Comparison of Computation Time for all Simulation Cases}
\label{subsubsection:ComparisonComputationTime}
\begin{figure}[htbp]
    \centering
    \includegraphics[width=0.5\textwidth]{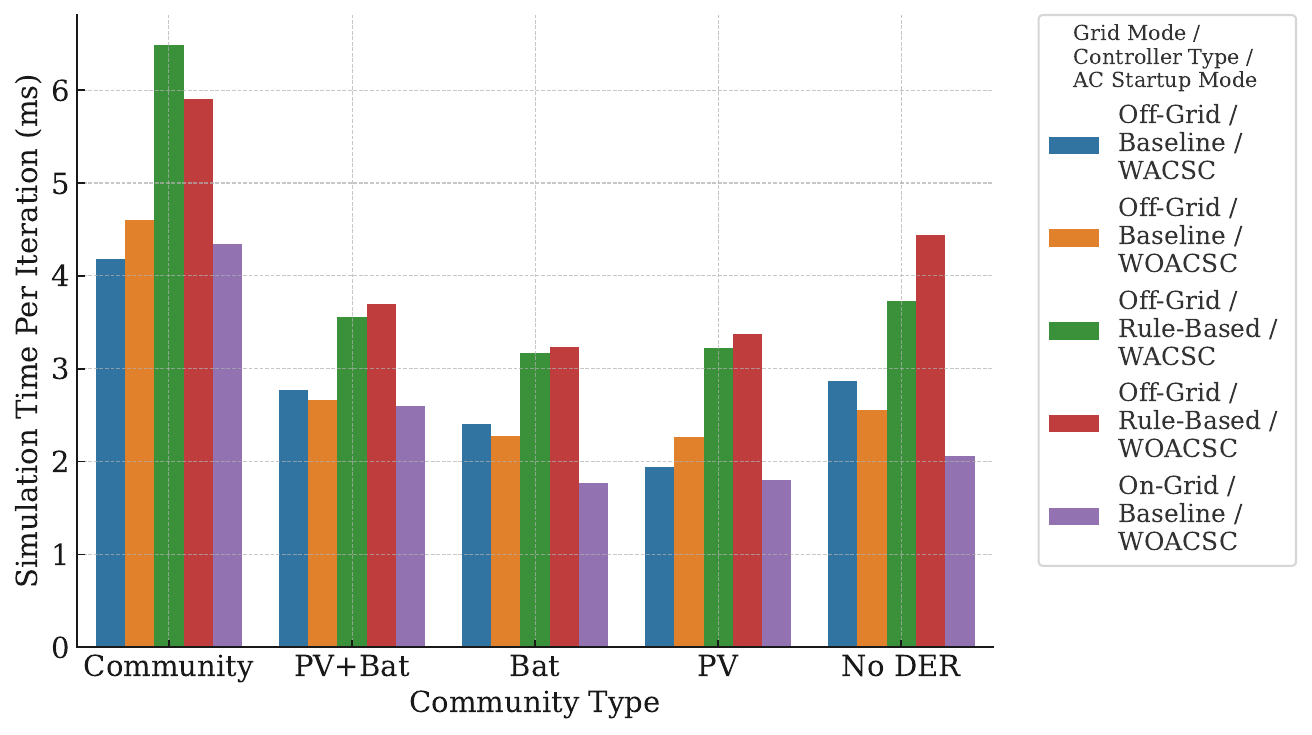}
    \caption{Comparison of performance metrics for all Community Types across all Grid Modes, Controller Types, and AC Startup Modes.}
    \label{fig:sim_time_per_iter}
\end{figure}

To evaluate the computational efficiency of the Smart Community Simulator, we compare the average simulation time per iteration across all community types and simulation types, including combinations of grid mode, controller type, and AC startup logic. Fig.~\ref{fig:sim_time_per_iter} presents these times in milliseconds for each case. These values account for the full simulation stack, including controller logic, energy balance physics, and thermal dynamics.

As shown, On-Grid Baseline (WOACSC) simulations exhibit the lowest computation times across all community types, due to the absence of AC startup constraints and minimal control overhead. For example, PV and PV+Bat configurations require less than 2 ms per iteration in this mode. In contrast, Off-Grid Rule-Based simulations—especially those using WACSC—have the highest runtimes, with the Community configuration reaching over 6 ms per iteration. This is expected, as these cases involve more frequent and complex decision-making, startup constraint checks, and multi-device coordination. 

It is important to note that these timings include the execution of legacy MATLAB-based components via the MATLAB Engine API. Once these modules are fully ported to Python, the total simulation time per iteration is expected to decrease significantly. Despite this, the current performance is already well within the bounds required for fast offline simulation, analysis, and closed-loop data-driven control development. These results highlight the simulator’s suitability as a lightweight and flexible platform for scalable testing, controller training, and algorithm benchmarking in smart grid applications.

\section{Conclusion}\label{section:Conclusion}
This paper introduced the Smart Community Simulator—an open-source, scalable, and modular simulation platform for developing, testing, and benchmarking Home Energy Management Systems (HEMS) at both the single-house and community levels. The simulator supports any number of houses with heterogeneous distributed energy resource (DER) configurations, real-world weather and load data pipelines, and both on-grid and off-grid modes. It is made available as a Gymnasium-compatible environment, enabling seamless integration with reinforcement learning and other control frameworks.

The case study demonstrated the simulator's flexibility across 25 diverse simulation cases, highlighting its ability to capture complex interactions between DERs, loads, and grid modes. The built-in controllers—Baseline and Rule-Based—serve as intuitive and lightweight benchmarks for control development. The built-in performance metrics enable an intuitive and quantitative assessment of thermal comfort, load resilience, and local generation sufficiency, supporting both simulation analysis and controller evaluation.

All simulation cases achieve sub-10 millisecond runtimes per iteration, even with the current MATLAB-Python hybrid implementation. These runtimes are suitable for fast offline simulation, analysis, and learning-based control development. Future versions of the simulator will include electric vehicles and hot water systems as controllable DERs, enhancing its relevance for emerging smart grid applications. Complete migration of the code base to Python will further improve computational performance and portability. This simulator will also serve as the foundation for our ongoing effort to develop a co-simulation platform integrated with OpenDSS (an electric power distribution system simulator) for testing and developing intelligent energy management algorithms under operational grid constraints.


\section*{Acknowledgment}
This work is funded by the NSF award 2208783.

\bibliographystyle{IEEEtran}
\input{output.bbl}

\end{document}